\documentclass[12pt]{article}
\usepackage{amssymb}
\usepackage[T1]{fontenc}
\usepackage[utf8]{inputenc}
\usepackage[intlimits]{amsmath}
\usepackage{enumerate}
\usepackage[dvips]{graphicx}
\usepackage[all]{xy}

\usepackage{bbm}

\newtheorem{lema}{Proposition}

\begin{document}
\begin{center}\Huge
Integrable Hamiltonian  systems related to the Hilbert--Schmidt ideal
\end{center}
\begin{center}\large
Anatol Odzijewicz and Alina Dobrogowska
\end{center}\begin{center}
Institute of Mathematics, University of Białystok\\
Akademicka 2, 15-267 Białystok, Poland
\end{center}
\begin{center}
E-mail: aodzijew@uwb.edu.pl, alaryzko@alpha.uwb.edu.pl
\end{center}
\begin{center}
{\bf Abstract}
\end{center}

By application of  the coinduction method as well as Magri method to the ideal of real Hilbert--Schmidt operators
 we construct the hierarchies
 of integrable  Hamiltonian  systems on the Banach Lie--Poisson spaces which consist of these type of operators.
  We also discuss their algebraic and analytic properties as well as solve them  in dimensions $N=2,3,4$.

\section{Introduction}
The concept of Banach Lie--Poisson space is the direct generalization of the
vector space with linear Poisson bracket to the context of Banach spaces
category. Namely, the Poisson bracket $\{f,g\}$ of smooth functions $f,g\in
C^{\infty}(\mathfrak{b})$ defined on Banach space $\mathfrak{b}$ is linear if
$\{\mathfrak{b}^{*}, \mathfrak{b}^{*}\}\subset\mathfrak{b}^{*}$, where
$\mathfrak{b}^{*}\subset C^{\infty}(\mathfrak{b})$ is dual space of
$\mathfrak{b}$. If one assumes additionaly that $\mathfrak{b}^{*}$ with
$[.,.]:=\{.,.\}|_{\mathfrak{b}^{*}}$ is a Banach Lie algebra $\mathfrak{g}$ such
that
\begin{equation}
\label{0}
ad_x^{*}\mathfrak{g}_{*}\subset\mathfrak{g}_{*},
\end{equation}
where $\mathfrak{g}_{*}:=\mathfrak{b}$, $x\in\mathfrak{g}$ and $ad_x:=[x,.]$, then we define on $\mathfrak{b}$ the structure of Banach Lie-Poisson space, see \cite{1}.

Equivalently, having Banach Lie algebra $\mathfrak{g}$ with predual $\mathfrak {g}_{*}$ which satisfies (\ref{0}) one defines linear Poisson bracket on $C^{\infty}(\mathfrak{g}_{*})$ by
\begin{equation}
\label{01}
\{f,g\}(b):=\left\langle b,[Df(b), Dg(b)]\right\rangle,
\end{equation}
where $b\in\mathfrak{b}$ and $f,g\in C^{\infty}(\mathfrak{g_{*}})$. The condition
(\ref{0}) is necessary for the existence of the Hamiltonian fields on
$\mathfrak{b}$. Since (\ref{0}) is automatically fulfilled if $\mathfrak{b}$ is
reflexive, this is the reason that one does not assume it in the finite
dimensional case.

In this paper we apply the methods of Banach--Poisson geometry elaborated in
\cite{1}, \cite{2} to the Banach Lie--Poisson spaces related to the ideal ${\cal
L}^2$ of  real Hilbert--Schmidt operators. Considering  various splittings of
${\cal L}^2$ and using the coinduction procedure (see \cite{2}) we obtain infinite
dimensional deformations of known hierarchies of integrable Hamiltonian
systems as well as other hierarchy which as we suppose is a new one.

The paper is organized as follows. In Section 2 we discuss various Banach Lie--Poisson spaces
obtained from the ideal of real Hilbert--Schmidt operators by the coinduction method suggested in
 \cite{2}. In particular, we present explicit formulas for Poisson brackets and the coadjoint actions
 for the Banach Lie groups and algebras taken under  considerations. We also show that Hamilton equations (\ref{a12}) on the Banach Lie--Poisson space ${\cal S}^2_{\alpha} $, with Hamiltonians (\ref{a2}) give  $k$ -- diagonal Toda lattice deformed by the bounded operator $A$ defined in (\ref{11}).

In Section 3 we obtain some hierarchy of Hamilton equations on the Banach
Lie-Poisson space $({\cal L}^2_{+},\{.,.\}_{+,\alpha})$ , see (\ref{a39}),  which is
parameterized by the operators $A$ defined in (\ref{11}). This new  hierarchy is
constructed by Magri method \cite{5} applied here in the context of Banach
Lie--Poisson  spaces theory.

We discuss equations (\ref{a39}) for special choice of $A$ in Section 4. Among
other we show that right hand side of equation (\ref{3d}) from the hierarchy
(\ref{a39}) can be realized by non-linear integral operator, see (\ref{e20eee}).
We also analize  (\ref{3d}) in details in the finite dimensional case and solve it
if dimension of the underlying Hilbert space is $N=2,3$ and $4$.

\section{Banach Lie--Poisson spaces \mbox{coinduced from ${\cal L}^2$}}

Let $\cal{H}$ be  the real separable Hilbert space  with  fixed orthonormal  basis \mbox{$\{\left. |
n\right\rangle \}_{n=0}^{\infty}$,}
i.e. $\left\langle n|m\right\rangle
=\delta_{nm}$. By ${\cal L}^{\infty}$ and ${\cal L}^{2}$ we denote the real Banach Lie algebra of bounded operators
and real Hilbert space of Hilbert--Schmidt operators acting in $\cal{H}$. We recall that ${\cal L}^{2}$ is an
ideal in ${\cal L}^{\infty}$ and there is  the  duality
$\left({\cal L}^{2}\right)^{*}\simeq {\cal L}^{2}$ defined by
\begin{equation}
\label{1}
\left\langle x,\rho\right\rangle :=Tr(x\rho),
\end{equation}
for $x,\rho\in {\cal L}^{2}$.
Note that paring (\ref{1}) is related to the scalar product of $x,y\in {\cal L}^2$
\begin{equation}
\left\langle x|y\right\rangle :=Tr(x^{\top}y),
\end{equation}
as follows
\begin{equation}
\left\langle x,\rho\right\rangle :=\left\langle x^{\top}|\;\rho\right\rangle
\end{equation}

According to (\ref{01}) one has the canonical Banach Lie-Poisson bracket on $C^{\infty}({\cal L}^2)$ given for
$f,g\in C^{\infty}({\cal L}^2)$ by
\begin{equation}
\label{2}
\{f,g\}(\rho):=Tr\left(\rho [Df(\rho),Dg(\rho)]\right).
\end{equation}


Let us expand the elements $\rho\in {\cal L}^2$ and $x\in {\cal L}^{\infty}$
\begin{align}
\label{4} & \rho=\sum_{n,m=0}^{\infty}\rho_{nm}\left. |n\right\rangle \left\langle m|\right. ,\\
\label{5} & x=\sum_{n,m=0}^{\infty}x_{nm}\left. |n\right\rangle \left\langle m|\right.
\end{align}
with respect to the orthonormal basis $\{\left. |n\right\rangle \left\langle m|\right.
\}_{n,m=0}^{\infty}$ of ${\cal L}^2$, where the series (\ref{4}) is convergent in
the $||.||_2$--norm and the series (\ref{5}) in the $w^*$--topology of ${\cal
L}^{\infty}$. In particular for the shift operator $S\in {\cal L}^{\infty}$ and its
adjoint $S^*\in {\cal L}^{\infty}$ one has the following decompositions
\begin{align}
\label{6} & S=\sum_{n=0}^{\infty}\left. |n\right\rangle \left\langle n+1|\right. ,\\
\label{7} & S^*=\sum_{n=0}^{\infty}\left. |n+1\right\rangle \left\langle n|\right. .
\end{align}
Using (\ref{4})  we define splitting
\begin{align}
\label{8} & {\cal L}^2={\cal L}^2_-\oplus {\cal L}^2_0\oplus {\cal L}^2_+
\end{align}
of  ${\cal L}^2$  into the sum of Hilbert subspaces of strictly  lower triangular
operators,  diagonal operators and strictly upper triangular operators
respectively. The pairing (\ref{1}) gives the following isomorphism of Hilbert
spaces:
\begin{equation}
\label{10}
\left({\cal L}^{2}_{-}\right)^{*}\cong {\cal L}^{2}_{+},\,\,\,\,\,
\left({\cal L}^{2}_{0}\right)^{*}\cong {\cal L}^{2}_{0},\,\,\,\,\,
\left({\cal L}^{2}_{+}\right)^{*}\cong {\cal L}^{2}_{-}.
\end{equation}

Let us take
\begin{equation}
 \label{10sss}
a:=\sum_{n=0}^{\infty}a_n\left. |n\right\rangle \left\langle n|\right.
\end{equation}
with
$||a||_{\infty}=\sup_{n\in\mathbb{N}\cup \{ 0\}}|a_n|\leq 1$, and define
the operator (not necessary bounded)
\begin{equation}
\label{11}
A:=\sum_{n=0}^{\infty}(aS)^n=\sum_{0\leq i\leq j}\alpha_{ij}\left. |i\right\rangle \left\langle j|\right. ,
\end{equation}
where
\begin{align}
\label{11a} \alpha_{ii}=1 & &\textrm{and} & &\alpha_{ij}=a_ia_{i+1}\ldots a_{j-1}  && \textrm{for} &&  i<j,
\end{align}
or equivalently
\begin{equation}
\label{11aa}
\alpha_{ij}\alpha_{jk}=\alpha_{ik}
\end{equation}
 if one puts $\alpha_{i,i+1}=a_i$.
Using operator $A$ we define the map
$\alpha:{\cal L}_{+}^{2}\rightarrow {\cal L}_{+}^{2}$ as follows
\begin{equation}
\label{12}
\alpha(x_{+}):=\sum_{0\leq i< j}\alpha_{ij}x_{ij}\left. |i\right\rangle \left\langle j|\right. ,
\end{equation}
where $x_+=\sum_{0\leq i<j}x_{ij}\left. |i\right\rangle \left\langle j|\right. \in {\cal L}_+^2$.

\begin{lema}
The map (\ref{12}) is a continuous endomorphism of the associative algebra ${\cal L}_+^{2}$
\begin{equation}
\label{12a}
\alpha(x_+y_+)=\alpha(x_+)\alpha(y_+),
\end{equation}
where $x_+,y_+\in {\cal L}_+^{2}$, if and only if (\ref{11aa}) is satisfied for
$0\leq i<j<k$.
\end{lema}
{\bf Proof.}
From (\ref{12}) by direct calculation we get
\begin{align}
\label{12aaa}
\alpha(x_+)\alpha(y_+)  & =
\left(\sum_{0\leq i< j}\alpha_{ij}x_{ij}\left. |i\right\rangle \left\langle j|\right.\right)
\left( \sum_{0\leq n< m}\alpha_{nm}y_{nm}\left. |n\right\rangle \left\langle m|\right.\right)=\\
&=\sum_{0\leq i< j}\sum_{0\leq n< m}\delta_{jn}\alpha_{ij}\alpha_{nm}x_{ij}y_{nm}\left. |i\right\rangle \left\langle m|\right.=\nonumber\\
& =\sum_{0\leq i< j<m}\alpha_{im}x_{ij}y_{jm}\left. |i\right\rangle \left\langle m|\right.=\alpha(x_+y_+).\nonumber
\end{align}
\hspace*{12cm}$\square$
\\

Taking the decompositions $\rho=\rho_-+\rho_0+\rho_+$ and
$x=x_-+x_0+x_+$ into account we define the Hilbert subspaces
\begin{align}
\label{13} & {\cal S}_{\alpha}^2:=\{\rho_{-}+\rho_0+\alpha\left(\rho_{-}^{\top}\right):
\rho_-\in {\cal L}^2_- \;\;\;\textrm{and}\;\;\; \rho_0\in {\cal L}^2_0\},\\
 & {\cal A}_{\alpha}^{2}:=\{x_--\alpha(x_{-}^{\top}): x_{-}\in {\cal L}^{2}_- \},\nonumber
\end{align}
considered as the components of the non-orthogonal  splittings
\begin{align}
\label{15} & {\cal L}^2={\cal S}^2_{\alpha}\oplus {\cal L}^2_+,\\
 & {\cal L}^{2}={\cal A}^{2}_{\alpha}\oplus {\cal L}^{2}_0\oplus {\cal L}^{2}_+.\nonumber
\end{align}

From (\ref{8}), (\ref{10}) and (\ref{15}) it follows that
\begin{align}
\label{17-17}  & \left({\cal S}^2_{\alpha}\right)^{*}\cong {\cal L}^{2}_{0} \oplus {\cal L}^{2}_+, & &
              \left({\cal L}^2_{+}\right)^{*}\cong {\cal A}^{2}_{\alpha}, \\
\label{17-18}   & \left({\cal L}^2_{+}\right)^{*}\cong {\cal L}^{2}_{-}, &&
          \left({\cal L}^2_-\oplus {\cal L}_0^2\right)^{*}\cong {\cal L}_0^{2}\oplus {\cal L}_+^{2}.
\end{align}

\begin{lema}
Hilbert subspaces ${\cal A}_{\alpha}^{2}$, ${\cal L}_0^{2}\oplus {\cal L}_+^{2}$  and ${\cal L}_-^{2}$  are Banach Lie subalgebras of $\left({\cal L}^{2}, [.,.]\right)$.\end{lema}
{\bf Proof.}
We prove that ${\cal A}_{\alpha}^{2}\subset {\cal L}^2$ is closed with respect to the commutator.
To this end we observe that operators
\begin{equation}
\label{19}
e_{ij}:=\left. |j\right\rangle \left\langle i|\right. -\alpha_{ij}\left. |i\right\rangle \left\langle j|\right. ,
\end{equation}
where $0\leq i< j$, form basis in ${\cal A}^{2}_{\alpha}$. Since
\begin{align}
\label{20}
[e_{ij},e_{nm}]= &  \delta_{mi}e_{nj} -\delta_{jn}e_{im}+\delta_{jm}
\left\{\begin{array}{ccc}
 -\alpha_{ij}e_{ni}  & for & n<i\\
 \alpha_{nj}e_{in} & for & i<n
\end{array}\right.+\nonumber\\
& +\delta_{in}
\left\{\begin{array}{ccc}
 -\alpha_{im}e_{mj}  & for & m<j\\
 \alpha_{ij}e_{jm} & for & j<m
\end{array}\right.,
\end{align}
we obtain $[{\cal A}^{2}_{\alpha},{\cal A}^{2}_{\alpha} ]\subset {\cal A}^{2}_{\alpha}$. The proof of inclusions
$[{\cal L}^{2}_{0}\oplus {\cal L}_+^{2},{\cal L}^{2}_{0}\oplus {\cal L}_+^{2} ]\subset {\cal L}^{2}_{0}\oplus {\cal L}_+^{2}$
 and $[{\cal L}^{2}_{-},{\cal L}^{2}_{-} ]\subset {\cal L}^{2}_{-}$ is staightforward.

\hspace*{12cm}$\square$
\\

The splittings ${\cal L}^2={\cal S}^2_{\alpha}\oplus {\cal L}_+^2$ and
${\cal L}^2=\left({\cal L}_-^2\oplus {\cal L}^2_0\right)\oplus  {\cal L}^2_{+} $ generate the
corresponding couples of projectors on their components
\begin{align}
\label{21} & \pi_{\alpha}:{\cal L}^2\rightarrow {\cal S}_{\alpha}^2,  & &
             \pi_{+}:{\cal L}^2\rightarrow {\cal L}_{+}^2, \\
          &  \pi_{+, \alpha}:{\cal L}^2\rightarrow {\cal L}_{+}^2, & &
              \pi_{-,0}:{\cal L}^2\rightarrow {\cal L}_{-}^2\oplus {\cal L}_0^2. & &
               & &  \nonumber
\end{align}
Let us denote by
\begin{align}
& \iota_{\alpha}:{\cal S}_{\alpha}^2 \hookrightarrow {\cal L}^2, &&
  \iota_{+}:{\cal L}_{+}^2 \hookrightarrow {\cal L}^2,\\
&  \iota_{+,\alpha}:{\cal L}_{+}^2  \hookrightarrow {\cal L}^2, &&
  \iota_{-,0}:{\cal L}_{-}^2\oplus {\cal L}_0^2 \hookrightarrow {\cal L}^2
\end{align}
the appropriate embeddings. The above notations are summarized in the
diagram

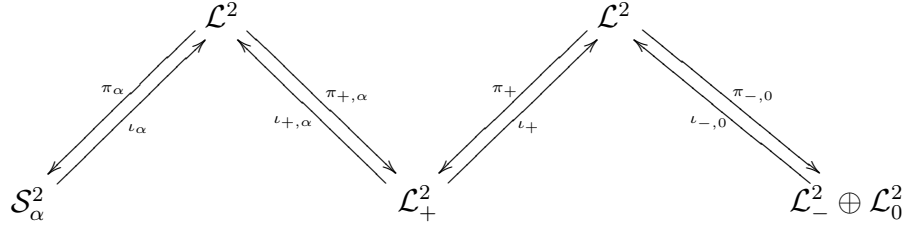
\hspace*{0.5cm}
\xymatrix{
                     & & {\cal L}^2 \ar@<.5ex>[ddrr]^*-<.9ex>\txt{\tiny{$\pi_{+,\alpha}$}}
             \ar@<-.5ex>[ddll]^*-<4ex>\txt{\tiny{$\pi_{\alpha}$}}      &&
             & & {\cal L}^2 \ar@<.5ex>[ddrr]^*-<.9ex>\txt{\tiny{$\pi_{-,0}$}}
             \ar@<-.5ex>[ddll]^*-<4ex>\txt{\tiny{$\pi_{+}$}} && \\
             & &&&    &&&&\\
{\cal S}_{\alpha}^2  \ar@<-.5ex>[uurr]^*-<4ex>\txt{\tiny{$\iota_{\alpha}$}}   &
&&     & {\cal L}_+^2
 \ar@<.5ex>[uull]^*-<.9ex>\txt{\tiny{$\iota_{+,\alpha}$}} \ar@<-0.5ex>[uurr]^*-<4ex>\txt{\tiny{$\iota_{+}$}}
 &
&& &     {\cal L}_-^2\oplus {\cal L}_0^2
 \ar@<.5ex>[uull]^*-<.9ex>\txt{\tiny{$\iota_{-,0}$}}
}

Let us note that the  isomorphisms (\ref{17-17}-\ref{17-18}) enable us to
consider derivatives $Df(\rho)$ and $Dg(\rho)$ as elements of the suitable
Banach--Lie subalgebras. Hence, one can  obtain immediately from the general
formula (\ref{01}) the Poisson brackets $\{f,g\}_{\alpha}$, $\{f,g\}_{-,0}$ on the
${\cal S}_{\alpha}^2$, ${\cal L}_{-}^2\oplus {\cal L}_0^2$ and
$\{f,g\}_{+,\alpha}$, $\{f,g\}_{+}$ on the ${\cal L}_{+}^2$,  respectively.
According to Proposition 2.4 in \cite{2} the following relations are valid
\begin{align}
\label{24} & \{f,g\}_{\alpha}=\{f\circ \pi_{\alpha}, g\circ \pi_{\alpha}\}\circ \iota_{\alpha}
& & \textrm{for} \;\; f,g\in C^{\infty}({\cal S}_{\alpha}^2), \\
\label{25} & \{f,g\}_{+,\alpha}=\{f\circ \pi_{+,\alpha}, g\circ \pi_{+,\alpha}\}\circ \iota_{+,\alpha}
& & \textrm{for} \;\; f,g\in C^{\infty}({\cal L}_{+}^2),\\
\label{26} & \{f,g\}_{+}=\{f\circ \pi_{+}, g\circ \pi_{+}\}\circ \iota_{+}
& & \textrm{for} \;\; f,g\in C^{\infty}({\cal L}_{+}^2),\\
\label{27} & \{f,g\}_{-,0}=\{f\circ \pi_{-,0}, g\circ \pi_{-,0}\}\circ \iota_{-,0}
& & \textrm{for} \;\; f,g\in C^{\infty}({\cal L}_{-}^2\oplus {\cal L}_0^2).
\end{align}
Since
\begin{equation}\iota_{\alpha}\circ \pi_{\alpha}=id,\;\;\;\iota_{+,\alpha}\circ
\pi_{+,\alpha}=id,\;\;\; \iota_{+}\circ \pi_{+}=id,\;\; \;\iota_{-,0}\circ \pi_{-,0}=id
\end{equation}
we conclude from (\ref{29}-\ref{27}) that projections $\pi_{\alpha}$,
$\pi_{+,\alpha}$, $\pi_+$ and $\pi_{-,0}$ are Poisson maps. So, the Poisson
brackets taken in (\ref{24}-\ref{27}) are coinduced by the bracket $\{.,.\}$
defined in (\ref{2}).

The maps  $\pi_{-,0}\circ \iota_{\alpha}:{\cal S}^2_{\alpha}\rightarrow {\cal
L}_{-}^2\oplus {\cal L}^2_{0}$ and $\pi_{\alpha}\circ \iota_{-,0}:  {\cal
L}_{-}^2\oplus {\cal L}^2_{0}\rightarrow {\cal S}^2_{\alpha}$ are mutually
inverse morphisms of Banach Lie--Poisson spaces. Taken in the coordinates
$\left( \rho_-,\rho_0\right)\in {\cal L}^2_-\oplus {\cal L}_0^2$ they assume
 the form of the identity map, i.e. $\pi_{-,0}\circ \iota_{\alpha}\left( \rho_-,\rho_0\right)=
\left( \rho_-,\rho_0\right)$. Thus in these coordinates the brackets (\ref{24}) and
(\ref{27})
 are given by
\begin{align}
\label{28}
\{f,g\}_{\alpha}(\rho_-,\rho_0) = \{f,g\}_{-,0}(\rho_-,\rho_0)
= &  -Tr
\left\{
\rho^{\top}_-\left(
\left[ \frac{\delta f}{\delta\rho_0}(\rho_-,\rho_0),\frac{\delta g}{\delta\rho_-}(\rho_-,\rho_0)\right]+\right.\right.\nonumber\\
& +\left[ \frac{\delta f}{\delta\rho_-}(\rho_-,\rho_0),\frac{\delta g}{\delta\rho_0}(\rho_-,\rho_0)\right]+  \\
& +\left.\left.\left[ \frac{\delta f}{\delta\rho_-}(\rho_-,\rho_0),\frac{\delta g}{\delta\rho_-}(\rho_-,\rho_0)\right]
\right)
\right\},
\nonumber
\end{align}
where $ \frac{\delta f}{\delta\rho_-}(\rho_-,\rho_0)$ and $\frac{\delta
f}{\delta\rho_0}(\rho_-,\rho_0)$ are the partial Fr\'echet derivatives
corresponding to $\rho_-\in {\cal L}^2_-$ and $\rho_0\in {\cal L}^2_0$. Let us
remark here that in coordinates $(\rho_{-},\rho_0)$  brackets (\ref{24}) and
(\ref{27}) do not depend on $\alpha$.

The explicit form of the bracket (\ref{25}) is the following
\begin{align}
\label{29}
\{f,g\}_{+, \alpha}(\rho_+)=  & Tr
\left\{
\rho_+
\left[\left( D f(\rho_+)\right)^{\top}-
\alpha\left( D f(\rho_+) \right),\right.\right.\\
& \left.\left.
\left( D g(\rho_+)\right)^{\top}-
\alpha\left( D g(\rho_+) \right)\right]
\right\},\nonumber
\end{align}
where  $D f(\rho_+)$  denotes the  Fr\'echet derivative in $\rho_+\in {\cal L}^2_+$.
One obtains the coordinate expression for the bracket (\ref{26})  from (\ref{29}) by putting
$\alpha_{ij}=0$ for $i<j$.

Using (\ref{29}) we get
\begin{align}
\label{a38} \{ \rho_{ij}, h\}_{+, \alpha}= &
 \sum_{n=i+1}^{j-1}
\left( \alpha_{nj}\rho_{in}\frac{\partial h}{\partial
\rho_{nj}}-
\alpha_{in}\frac{\partial h}{\partial \rho_{in}}\rho_{nj}\right)+\\
& +\sum_{n=0}^{i-1} \left( \rho_{nj}\frac{\partial h}{\partial
\rho_{ni}}-
\alpha_{ij}\frac{\partial h}{\partial \rho_{nj}}\rho_{ni}\right)+\nonumber\\
& +\sum_{n=j+1}^{\infty} \left( \alpha_{ij}\rho_{jn}\frac{\partial
h}{\partial \rho_{in}}- \frac{\partial h}{\partial
\rho_{jn}}\rho_{in}\right),\nonumber
\end{align}
where $h\in C^{\infty}({\cal L}^2_{+})$ and $\rho_{ij}$, with $i<j$ and $i,j \in\mathbb{N}\cup\{ 0\}$,
are coordinate functions of   $\rho_+\in {\cal L}^2_+$ given by (\ref{4}).
In particular case we have the Poisson bracket
\begin{align}
\label{20aaa}
\{ \rho_{ij},\rho_{nm}\}_{+,\alpha}= &  \delta_{mi}\rho_{nj} -\delta_{jn}\rho_{im}+\delta_{jm}
\left\{\begin{array}{ccc}
 -\alpha_{ij}\rho_{ni}  & \textrm{for} & n<i\\
 \alpha_{nj}\rho_{in} & \textrm{for} & i<n
\end{array}\right.+\nonumber\\
& +\delta_{in}
\left\{\begin{array}{ccc}
 -\alpha_{im}\rho_{mj}  & \textrm{for} & m<j\\
 \alpha_{ij}\rho_{jm} & \textrm{for} & j<m
\end{array}\right.
\end{align}
between the coordinate functions.

 The Lie algebra ${\cal L}^2$ is Banach--Lie algebra of Banach--Lie group $G{\cal L}^2$
 which consists of  invertible bounded operators of the form  $1+x\in {\cal L}^{\infty}$,
 where $x\in {\cal L}^2$. Since $1+x\in G{\cal L}^{2}$
for $||x||_2<1$ one can identify $G{\cal L}^2$  with the open subset of ${\cal
L}^2$. One has the following  Banach--Lie subgroups
\begin{align}
& G\left( {\cal L}_0^{2}\oplus {\cal L}_+^{2}\right)=\{ 1+x\in G{\cal L}^{2}:x\in{\cal L}_0^{2}\oplus {\cal L}_+^{2}\},\\
& G{\cal L}_{-}^{2}=\{ 1+x\in G{\cal L}^{2}:x\in {\cal L}_{-}^{2}\}
\end{align}
which have  ${\cal L}_0^{2}\oplus {\cal L}_+^{2}$ and ${\cal L}_{-}^{2}$ as their
Banach--Lie algebras. We define the Banach--Lie subgroup $G{\cal
A}_{\alpha}^{2}\subset G{\cal L}^{2} $ as follows
\begin{equation}
 G{\cal A}_{\alpha}^{2}:=\{exp\; x\in G{\cal L}^{2}: x\in {\cal A}_{\alpha}^{2}\}.
\end{equation}
Let us recall that $x\in {\cal A}_{\alpha}^{2}$ iff
\begin{equation}
 x=\sum_{0\leq i<j}x_{ij}e_{ij},
\end{equation}
where the basis $\{e_{ij}\}$ is defined in (\ref{19}).
In Section 3 we will back to the investigation of these groups.

Now we present the formulas for the coadjoint actions of the Banach--Lie group
$G\left( {\cal L}_0^{2}\oplus {\cal L}_+^{2}\right)$ on ${\cal S}_{\alpha}^2$ and on ${\cal L}^2_-\oplus {\cal L}^2_0$. We have
\begin{align}
\label{a6} \left( Ad^{\alpha}\right)^*_{g^{-1}}\rho & = \pi_{\alpha}\left(g\iota_{\alpha}(\rho)g^{-1}\right)=\\
& =
\pi_{-,0}\left(g\iota_{\alpha}(\rho)g^{-1}\right)+\alpha\left(\pi_+\left(\left(g\iota_{\alpha}(\rho)g^{-1}\right)^{\top}\right)\right),\nonumber
\end{align}
where $\rho\in {\cal S}_{\alpha}^2$, $  g\in G\left( {\cal L}_0^{2}\oplus {\cal L}_+^{2}\right)$, and
\begin{align}
\label{a7} \left( Ad^{-,0}\right)^*_{g^{-1}}\rho & = \pi_{-,0}\left(g\iota_{-,0}(\rho)g^{-1}\right),
\end{align}
where $\rho\in {\cal L}^2_-\oplus {\cal L}^2_0$, $ g\in G\left( {\cal L}_0^{2}\oplus {\cal L}_+^{2}\right)$, respectively.

The coadjoint action of the Banach--Lie group $G{\cal A}^{2}_{\alpha}$ on ${\cal L}^2_+$ is as follows
\begin{align}
\label{a8}  \left( Ad^{+,\alpha}\right)^*_{g^{-1}}\rho & =   \pi_{+,\alpha}\left(g\iota_{+,\alpha}(\rho)g^{-1}\right)= \\
 &=
\pi_+\left(g\iota_{+,\alpha}(\rho)g^{-1}\right)-\alpha\left(\pi_+\left(\left(g\iota_{+,\alpha}(\rho)g^{-1}\right)^{\top}\right)\right),\nonumber
\end{align}
where $ \rho\in {\cal L}^2_+$, $ g\in G{\cal A}^{2}_{\alpha}$.

The coadjoint actions $\left( ad^{\alpha}\right)^*$, $\left( ad^{-, 0}\right)^*$, $\left( ad^{+,\alpha}\right)^*$
of the corresponding Banach--Lie algebras  on the related Banach
Lie--Poisson spaces are given by
\begin{align}
\label{a9} & \left( ad^{\alpha}\right)^*_{x}\rho=-\pi_{\alpha}\left([x,\iota_{\alpha}(\rho)]\right),
\end{align}
for $\rho\in {\cal S}_{\alpha}^2$,  $x\in {\cal L}_0^{2}\oplus {\cal L}_+^{2}$,
\begin{align}
\label{a10} & \left( ad^{-,0}\right)^*_{x}\rho=-\pi_{-,0}\left([x,\iota_{-,0}(\rho)]\right)
\end{align}
for $\rho\in {\cal L}^2_-\oplus {\cal L}^2_0$,  $x\in  {\cal L}_0^{2}\oplus {\cal L}_+^{2}$,
\begin{align}
\label{a11} & \left( ad^{+,\alpha}\right)^*_{x}\rho=-\pi_{+,\alpha}\left([x,\iota_{+,\alpha}(\rho)]\right)
\end{align}
for $\rho\in {\cal L}^2_+$, $x\in {\cal A}^{2}_{\alpha}$. From  (\ref{a9}), (\ref{a10}) and (\ref{a11}) we obtain the Hamilton equations
\begin{align}
\label{a12}  \frac{d}{dt}(\rho_-+\rho_0) & = -\left( ad^{\alpha}\right)^*_{Dh(\rho_-,\rho_0)}(\rho_-+\rho_0)=\\
               &=\pi_{\alpha}\left([Dh(\rho_-,\rho_0),\iota_{\alpha}(\rho_-+\rho_0)]\right),\nonumber\\
\label{a12a}            \frac{d}{dt}(\rho_-+\rho_0) & =-\left( ad^{-,0}\right)^*_{Dh(\rho_-,\rho_0)}(\rho_-+\rho_0)=\\
     & = \pi_{-,0}\left([Dh(\rho_-,\rho_0),\iota_{-,0}(\rho_-+\rho_0)]\right),\nonumber\\
\label{a13}  \frac{d}{dt}\rho_+ & = -\left( ad^{+,\alpha}\right)^*_{Dh(\rho_+)}\rho_+=
            \pi_{+,\alpha}\left([Dh(\rho_+),\iota_{+,\alpha}(\rho_+)]\right)
\end{align}
on the Banach Lie-Poisson spaces ${\cal S}^2_{\alpha}$, ${\cal L}_-^{2}\oplus {\cal L}_0^{2}$  and ${\cal L}^2_+$ associated with the Hamiltonians $h\in C^{\infty}\left({\cal S}^2_{\alpha}\right)$, $h\in C^{\infty}\left( {\cal L}_-^{2}\oplus {\cal L}_0^{2}\right)$  and
$h\in C^{\infty}\left({\cal L}^2_{+}\right)$, respectively.

Using the methods elaborated in \cite{2} one shows that the   Casimirs
\begin{align}
\label{a1} & I_l(\rho):=\frac{1}{l} Tr \rho^l,\\
           &\{I_l,I_k\}=0,\;\;\;\;\;1<l\in\mathbb{N},\nonumber
\end{align}
defined on Banach Lie--Poisson space $({\cal L}^2,\{.,.\})$ after restriction to
${\cal S}_{\alpha}^2$ and ${\cal L}_-^2\oplus {\cal L}^2_0$   give  infinite  families of   integrals in involution:
\begin{align}
\label{a2} & I_l^{\alpha}(\rho_-,\rho_0):=\frac{1}{l} Tr \left(\rho_-+\rho_0+\alpha(\rho^{\top}_-)\right)^l,\\
           &\{I_l^{\alpha},I_k^{\alpha}\}_{\alpha}=0,\nonumber\\
       & \nonumber \\
\label{a4} & I_l^{-,0}(\rho_-,\rho_0):=\frac{1}{l} Tr \left(\rho_-+\rho_0\right)^l=\frac{1}{l} Tr \rho_0^l,\\
           &\{I_l^{-,0},I_k^{-,0}\}_{-,0}=0,\nonumber
\end{align}
on the isomorphic Banach Lie--Poisson spaces
\begin{equation}
 \label{aa2}
\left({\cal S}^2_{\alpha}, \{.,.\}_{\alpha}\right)\cong
\left({\cal L}^2_{-}\oplus {\cal L}^2_0, \{.,.\}_{-,0}\right).
\end{equation}
Hamilton equations in the case (\ref{a4}) are easy to integrate. The case
(\ref{a2}) is the most interesting from the point of view  of the theory of
integrable Hamiltonian systems. In order to clarify this we use the isomorphism
(\ref{aa2}) and consider Banach spaces embeddings $\iota _k:{\cal L}^2_{-,k}
\hookrightarrow {\cal L}^1_{-}\oplus {\cal L}^2_{0}$, where $\varrho\in {\cal
L}^2_{-,k} \subset {\cal L}^2_-\oplus  {\cal L}^2_0$ iff
\begin{equation}
\label{g2}
\varrho=\sum_{i=0}^{k-1}\left(S^*\right)^i\varrho_i,\;\;\; \varrho_i\in {\cal L}^2_0.
\end{equation}
According to \cite{2}, ${\cal L}^2_{-,k}$ is a Banach Lie-Poisson space with the
Poisson bracket of $f,g \in C^{\infty}\left({\cal L}^2_{-,k}\right)$ given by
\begin{equation}
\label{g1}
\{ f, g\}_{-,k}=\sum_{l=0}^{k-1}\sum_{i=0}^{l} Tr \left[ \varrho_l\left(\frac{\delta f}{\delta\varrho_i}(\varrho)s^i
\left( \frac{\delta g}{\delta\varrho_{l-1}}(\varrho)\right)-\right.\right.
\end{equation}$$-\left.\left.
\frac{\delta g}{\delta\varrho_i}(\varrho)s^i
\left( \frac{\delta f}{\delta\varrho_{l-1}}(\varrho)\right)
\right)\right],
$$
where $\frac{\delta f}{\delta \varrho_i}(\varrho)$ denotes the partial  derivative of $f$ in $\varrho_{i}$ defined by the expansion
\begin{equation}
Df(\varrho)=\frac{\delta f}{\delta \varrho_0}(\varrho)+\frac{\delta f}{\delta \varrho_1}(\varrho) S+\ldots +
\frac{\delta f}{\delta \varrho_{k-1}}(\varrho) S^{k-1}
\end{equation}
and we used the following notation
\begin{equation}
s^i\left( \sum_{n=0}^{\infty}x_n\left| n\right\rangle\left\langle n\right|\right):=\sum_{n=0}^{\infty}x_{n+1}\left| n\right\rangle\left\langle n\right|.
\end{equation}
The embedding $\iota _k:{\cal L}^2_{-,k}
\hookrightarrow {\cal L}^2_{-}\oplus {\cal L}^2_{0}$ is a Poisson map. Thus $I^{\alpha}_{l}\circ\iota_k$ form on ${\cal L}^2_{-,k}$ an infinite hierarchy of integrals in the involution
\begin{equation}
\{  I^{\alpha}_{l}\circ\iota_k, I^{\alpha}_{m}\circ\iota_k \}_{-,k}=0,\;\;\; l,m\in \mathbb{N}.
\end{equation}
The corresponding hierarchy of Hamilton equations is as follows
\begin{equation}
\frac{\partial \varrho_j}{\partial t_l} =-\sum_{l=j}^{k-1}
\left(\widetilde{s}^{l-j}
\left(\varrho_l\frac{I^{\alpha}_{l}\circ\iota_k}
{\delta\varrho_{l-j}}(\varrho)\right)-\varrho_l s^j\left(\frac{I^{\alpha}_{l}\circ\iota_k}
{\delta\varrho_{l-j}}(\varrho)\right)
\right),
\end{equation}
where $\varrho_j\in {\cal L}^2_0$, $j=0,1,\ldots ,k-1$, are given by (\ref{g2}) and
\begin{equation}
\widetilde{s}^{\;l}\left( \sum_{n=0}^{\infty}x_n\left| n\right\rangle\left\langle n\right|\right):=\sum_{n=0}^{\infty}x_{n}\left| n+l\right\rangle\left\langle n+l\right|.
\end{equation}
For $A={\mathbbm 1}$, i.e.
$\alpha_{ii}=1$ and $\alpha_{ij}=0$ if $i<j$, the system of integrals in involution (\ref{a2}) reduces
 to the one given by (\ref{a4}).
For the case $A=\sum_{n=0}^{\infty}S^n$, i.e. $\alpha_{ij}\equiv 1$, the integrals of motion
(\ref{a2}) substituted in (\ref{a12}) generate the $k$--diagonal Toda hierarchies, see \cite{2}.

So, (\ref{a2}) gives an infinite parameter deformations of the $k$--diagonal
Toda systems. These deformations  with the exception of the generic case
(given by $\alpha_{ij}\neq 0)$ also include the singular subcases
corresponding to $a_i=0$ for some $i\in\mathbb{N}\cup \{0 \}$.

Restricting the Casimirs of $({\cal L}^2,\{.,.\})$ to $({\cal L}^2_+,\{.,.\}_+)$ and
$({\cal L}^2_+,\{.,.\}_{+,\alpha})$ we obtain the trivial
$I^+_l(\rho_+)=I^{+,\alpha}_l(\rho_+)\equiv0$ integrals of motion. Hence, for the
Banach Lie--Poisson spaces $({\cal L}^2_{+},\{.,.\}_{+,\alpha})$ one needs
different methods then the ones described above.

\section{Hamiltonian systems on Banach Lie--Poisson spaces $({\cal L}^2_{+},\{.,.\}_{+,\alpha})$}

Now we shall investigate  Banach--Lie algebras ${\cal A}^2_{\alpha}$ and the
corresponding Banach Lie-Poisson spaces $({\cal L}^2_{+},\{.,.\}_{+,\alpha})$
and the Banach--Lie groups $G{\cal A}^2_{\alpha}$ for the different values of
the deformation parameters $a_k\in{\mathbb R}$, $k\in{\mathbb N}\cup \{0\}$.
In our considerations we will  to some extent imitate the methods used in the
finite dimensional case, e.g. see \cite{7} and others therein, keeping in mind
however that not all finite dimensional constructions can be transferred to
Hilbert--Schmidt case without some additional conditions on  $a_k\in{\mathbb
R}$, $k\in{\mathbb N}\cup \{0\}$. Henceforth, we assume that
$\sum_{k=0}^{\infty}|1-a_k|<\infty$ and $a_k= 0$ for at the most  the finite
number of indexes $k_1<k_2\ldots <k_N$.

Let us define the following sequence of Lie subalgebras
\begin{equation}
\label{k1}
{\cal A}^2_{\alpha_l}:=\left\{\sum_{k_{l-1}\leq i<j\leq k_l}x_{ij} e_{ij}:\;\; x_{ij}\in{\mathbb R}\right\},
\end{equation}
where $k_0:=0$, $k_{N+1}:=\infty$ and $l=1,\ldots , N+1$. They can be included in the exact sequence of Banach--Lie algebras
\begin{equation}
 \label{k2}
0\stackrel{}{\longrightarrow}{\cal N}^2_{k_1,\ldots, k_N}\stackrel{\iota}{\longrightarrow}{\cal A}^2_{\alpha}\stackrel{\pi}{\longrightarrow}
\bigoplus^{N+1}_{l=1}{\cal A}^2_{\alpha_l}\stackrel{}{\longrightarrow} 0,
\end{equation}
for which the Lie ideal ${\cal N}^2_{k_1,\ldots ,k_N}$ is defined by
\begin{equation}
\label{k3}
{\cal N}^2_{k_1,\ldots, k_N}:=\left\{\sum_{l=1}^{N}\sum_{k_{l-1}\leq i\leq k_l<j}x_{ij} \left |j\right\rangle\left\langle i\right |:\;\; x_{ij}\in{\mathbb R}\right\}    .
\end{equation}
The exact sequence of Banach Lie--Poisson spaces predual to the one given in (\ref{k2})
 is the following
\begin{equation}
 \label{k4}
0\stackrel{}{\longrightarrow}\bigoplus^{N+1}_{l=1}{\cal L}^2_{+,k_l}
\stackrel{\pi_*}{\longrightarrow}{\cal L}^2_{+}\stackrel{\iota_*}{\longrightarrow}
{\cal L}^2_{+,k_1,\ldots , k_N}\stackrel{}{\longrightarrow} 0,
\end{equation}
where
\begin{equation}
 \label{k5}
{\cal L}^2_{+,k_l}:=\left\{\sum_{k_{l-1}\leq i<j\leq k_l}\rho_{ij} \left |i\right\rangle\left\langle j\right |:\;\; \rho_{ij}\in{\mathbb R}\right\},
\end{equation}
and
\begin{equation}
 \label{k6}
{\cal L}^2_{+,k_1,\ldots , k_N}:=\left\{\sum_{l=1}^{N}\sum_{k_{l-1}\leq i\leq k_l<j}\rho_{ij} \left |i\right\rangle\left\langle j\right |:\;\; \rho_{ij}\in{\mathbb R}\right\}   .
\end{equation}
Note here that the isomorphisms:
\begin{align}
 \label{k7} &\left(\bigoplus^{N+1}_{l=1}{\cal L}^2_{+,k_l}\right)^*\cong \bigoplus^{N+1}_{l=1}{\cal A}^2_{\alpha_l},
&  & \left({\cal L}^2_+\right)^*\cong {\cal A}^2_{\alpha} &
\end{align}
and
\begin{align}
 \label{k8}
& \left( {\cal L}^2_{+,k_1,\ldots , k_N}\right)^*\cong {\cal N}^2_{k_1,\ldots, k_N}&
\end{align}
take place. The exactness of sequences (\ref{k2}) and (\ref{k4}), where $\iota$,
$\pi$ are Banach--Lie algebras morphisms and their preduals $\pi_*$, $\iota_*$
are Poisson maps, respectively, is a consequence of Hilbert spaces splitings
\begin{align}
 \label{k9} & {\cal L}^2_+={\cal L}^2_{+,k_1,\ldots , k_N}\oplus\left( \bigoplus^{N+1}_{l=1}{\cal L}^2_{+,k_l}\right),\\
\label{k10} & {\cal A}^2_{\alpha}={\cal N}^2_{k_1,\ldots, k_N}\oplus\left(\bigoplus^{N+1}_{l=1}{\cal A}^2_{\alpha_l}\right)
\end{align}
into the orthogonal Hilbert subspaces.

Let us define two bounded diagonal operators
\begin{align}
&\label{a4ab} \eta_{\alpha}:=\sum_{i=0}^{\infty}\alpha_{i\infty}
|i\rangle\langle i|,\\
& \delta_{\alpha}:=\sum_{i=0}^{\infty}\alpha_{0i} |i\rangle\langle
i|,
\end{align}
where $ \alpha_{i\infty}:=\prod_{j=i}^{\infty}a_j$, which satisfy
\begin {equation}
\label{p2}
\eta_{\alpha}\delta_{\alpha}=\alpha_{0\infty}.
\end {equation}

For nonsingular case $\alpha_{0\infty}\neq 0$ the following isomorphisms  of Banach--Lie algebras
\begin{equation}
\label{p3}
({\cal A}^2_{\alpha},[.,.])\cong ({\cal O}( \eta_{\alpha}), [.,.])\cong ({\cal O}, [.,.]_{\eta_{\alpha}}),
\end{equation}
take place, where
\begin{equation}
 \label{p4}
{\cal O}( \eta_{\alpha}):=\{X\in{\cal L}^2:\;\; X\eta_{\alpha}+\eta_{\alpha}X^{\top}=0\}
\end{equation}
and
\begin{equation}
 \label{p5}
{\cal O}:=\{X\in{\cal L}^2:\;\; X+X^{\top}=0\}.
\end{equation}
The bracket $[X,Y]_{\eta_{\alpha}}$ of $X,Y\in{\cal O}$ is defined by
\begin{equation}
 \label{p6}
[X,Y]_{\eta_{\alpha}}:=X\eta_{\alpha}Y -Y\eta_{\alpha}X.
\end{equation}
The first isomorphism in (\ref{p3}) is due to the observation that
$X=x_{-}-\alpha (x_{-}^{\top})\in {\cal A}^2_{\alpha}$ is a solution of
$X\eta_{\alpha}+\eta_{\alpha}X^{\top}=0$ if $\alpha(x_{-}^{\top})=\eta_{\alpha} x_{-}^{\top}\eta_{\alpha}^{-1}$.
If $\alpha_{0\infty}\neq 0$ the map
\begin{equation}
\label{p7}
\phi_{\alpha}:{\cal O}\ni X\rightarrow \eta_{\alpha} X\in {\cal O}( \eta_{\alpha})
\end{equation}
is invertible continuous operator which define the second isomorphism  in
(\ref{p3}).

If $a_k=1$ and $a_k=0$, for all $k\in{\mathbb N}\cup \{0\}$, then one obtains
${\cal O}(\eta_{\alpha})={\cal O}$ and ${\cal O}(\eta_{\alpha})={\cal L}^2_{-}$
respectively. Two Banach--Lie algebras $\left({\cal
O},[.,.]_{\eta_{\alpha}}\right)$ and $\left({\cal O},[.,.]_{\eta_{\beta}}\right)$ are
isomorphic if operators $\eta_{\alpha}$ and $\eta_{\beta}$ have the same
signatures, i.e. if their sets of positive, negative and zero diagonal elements
have the same cardinalities.

In the singular case $\alpha_{0\infty}=0$ the Banach--Lie algebra
isomorphisms given by (\ref{p3}) are not valid in general. For example if
$a_0=0$ and $a_k=1$, for $k\in{\mathbb N}$, the Banach--Lie algebras
$\left({\cal A}^2_{\alpha},[.,.]\right)$ and $\left({\cal O},[.,.]_{\eta_{\alpha}}\right)$
are isomorphic to the Euclidian algebra ${\cal E}(\left|0\right\rangle^{\bot})$ of
the real Hilbert subspace $\left|0\right\rangle^{\bot}\subset {\cal H}$ orthogonal
to the first element of the basis fixed in Section 2. However the Banach--Lie
algebra $\left({\cal O}(\eta_{\alpha}),[.,.]\right)$ is not isomorphic to ${\cal
E}(\left|0\right\rangle^{\bot})$. When  $a_l=0$, for some $l>0$, and $a_k=1$,
for $k\neq l$, all Banach--Lie algebras in (\ref{p3}) are not isomorphic.

The Hilbert subspace ${\cal L}^2_+\subset{\cal L}^2$ of upper triangular
operators is predual to $({\cal A}^2_{\alpha},[.,.])$ as well as to $({\cal O},
[.,.]_{\eta_{\alpha}})$. Thus apart of the Poisson bracket $\{.,.\}_{+,\alpha}$ on
${\cal L}^2_+$ we have the second Poisson bracket
\begin{align}
 \label{p8}
\{ f,g\}_{\eta_{\alpha}}(\rho_+):= & Tr\left\{\rho_+\left[
\left(\frac{\delta f}{\delta\rho_+}(\rho_+)\right)^{\top}-\left(\frac{\delta f}{\delta\rho_+}(\rho_+)\right),\right.\right.
\\
& \left.\left.
\left(\frac{\delta g}{\delta\rho_+}(\rho_+)\right)^{\top}-\left(\frac{\delta g}{\delta\rho_+}(\rho_+)\right)\right]_{\eta_{\alpha}}\right\}\nonumber
\end{align}
for $f,g\in C^{\infty}({\cal L}^2_+)$.

 In the nonsingular case the map $R_{\eta_{\alpha}}:{\cal
L}^2_+\longrightarrow {\cal L}^2_+$ defined by
\begin{equation}
 \label{p9}
R_{\eta_{\alpha}}\rho_+:=\rho_+\eta_{\alpha},
\end{equation}
gives linear Poisson isomorphism between Banach Lie--Poisson spaces  $({\cal L}^2_+,$ $\{.,.\}_{+,\alpha})$ and $({\cal L}^2_+,\{.,.\}_{\eta_{\alpha}})$, i.e.
\begin{equation}
 \label{p10}
\{f\circ R_{\eta_{\alpha}},g\circ R_{\eta_{\alpha}}\}_{+,\alpha}=\{f,g\}_{\eta_{\alpha}}\circ R_{\eta_{\alpha}}.
\end{equation}

The complete systems of integrals in involution for the Lie--Poisson space
$({\cal L}^2_+,\{.,.\}_{\eta_{\alpha}})$, in finite dimensional case was described
by A. Bolsinov \cite{7}. In nonsingular case, since we have the isomorphism
(\ref{p10}), one obtains also the complete systems of integrals for $({\cal
L}^2_+,\{.,.\}_{+,\alpha})$. However, in order to find the integrals in singular
case one needs to apply different methods.

For $g\in G{\cal A}^2_{\alpha}$ one has relations
\begin{align}
\label{p1}
g\eta_{\alpha}g^{\top}=\eta_{\alpha},
&& g^{\top}\delta_{\alpha}g=\delta_{\alpha},
\end{align}
which allow us to find invariants of the coadjoint representation of the group  $G{\cal A}^2_{\alpha}$.

\begin{lema}
For $I^k_{\alpha}:{\cal L}^2_+\longrightarrow \mathbb{R}$ defined by
\begin{equation}
\label{a30} I^k_{\alpha}(\rho_+):= Tr
\left(\alpha_{0\infty}\rho_+^2-\rho_+\eta_{\alpha}\rho_+^{\top}\delta_{\alpha}-
\eta_{\alpha}\rho_+^{\top}\delta_{\alpha}\rho_++\eta_{\alpha}\left(\rho_+^{\top}\right)^2\delta_{\alpha}
\right)^{k}
\end{equation}
one has
\begin{equation}
\label{a31} I^k_{\alpha}\left( \left(
Ad^{+,\alpha}\right)^*_{g^{-1}}
\rho_+\right)=I^k_{\alpha}\left(\rho_+\right),
\end{equation}
where $g\in G{\cal A}^2_{\alpha}$.
\end{lema}
{\bf Proof.}
Substituting $\left( Ad^{+,\alpha}\right)^*_{g^{-1}}\rho_+$  given by (\ref{a8}) into (\ref{a30})
instead of $\rho_+$  from the identities
\begin{align}
 \eta_{\alpha}\left( \left(
Ad^{+,\alpha}\right)^*_{g^{-1}}\rho_+\right)^{\top}\delta_{\alpha}= &
\eta_{\alpha}\left(\pi_+(g\rho_+g^{-1})\right)^{\top}\delta_{\alpha}-\\
& -\alpha_{0\infty}\pi_+(g\rho_+g^{-1})^{\top},\nonumber\\
\eta_{\alpha}\left(\alpha(\rho_+)\right)^{\top}\delta_{\alpha}= & \alpha_{0\infty}\rho_+^{\top}, \\
 \eta_{\alpha}\rho_+\delta_{\alpha}= &\alpha_{0\infty}\alpha(\rho_+)
\end{align}
we get
\begin{align}
\label{x1}
& I^k_{\alpha}\left( \left( Ad^{+,\alpha}\right)^*_{g^{-1}}\rho_+\right)=
Tr\left(\alpha_{0\infty}\left(\pi_+(g\rho_+g^{-1})-\alpha\left(\pi_+(g\rho_+g^{-1})^{\top}\right)\right)^{2}-\right. \nonumber\\
& -\left(\pi_+(g\rho_+g^{-1})-\alpha\left(\pi_+(g\rho_+g^{-1})^{\top}\right)\right)
\nonumber\\
&\left(\eta_{\alpha}\left(\pi_+(g\rho_+g^{-1})\right)^{\top}\delta_{\alpha}-\alpha_{0\infty}
\pi_+(g\rho_+g^{-1})^{\top}
\right)-\nonumber\\
& -\left(\eta_{\alpha}\left(\pi_+(g\rho_+g^{-1})\right)^{\top}\delta_{\alpha}-\alpha_{0\infty}
\pi_+(g\rho_+g^{-1})^{\top}
\right)\nonumber\\
&\left(\pi_+(g\rho_+g^{-1})-\alpha\left(\pi_+(g\rho_+g^{-1})^{\top}\right)\right)+\nonumber \\
&\left. +\eta_{\alpha}
\left(\pi_+(g\rho_+g^{-1})-\alpha\left(\pi_+(g\rho_+g^{-1})^{\top}\right)\right)^{2\top}\delta_{\alpha}
\right)^k=\nonumber\\
& = Tr\left(\alpha_{0\infty}\left((\pi_+(g\rho_+g^{-1}))^2+\pi_+(g\rho_+g^{-1})\pi_+(g\rho_+g^{-1})^{\top}+ \right.\right.\\
& \left.+\pi_+(g\rho_+g^{-1})^{\top}\pi_+(g\rho_+g^{-1})+(\pi_-(g\rho_+g^{-1})^{\top})\right)^2-\nonumber\\
&-
\left(\pi_+(g\rho_+g^{-1})+\pi_+(g\rho_+g^{-1})^{\top}\right)
\eta_{\alpha}
\left((\pi_+(g\rho_+g^{-1}))^{\top}+\right.\nonumber\\
&\left. +(\pi_+(g\rho_+g^{-1})^{\top})^{\top}\right)\delta_{\alpha}-
\eta_{\alpha}\left((\pi_+(g\rho_+g^{-1}))^{\top}+\right.
\nonumber\\
& \left.
+(\pi_+(g\rho_+g^{-1})^{\top})^{\top}\right)\delta_{\alpha}
\left(\pi_+(g\rho_+g^{-1})+\pi_+(g\rho_+g^{-1})^{\top}\right)+\nonumber\\
& + \eta_{\alpha}
\left((\pi_+(g\rho_+g^{-1}))^2+\pi_+(g\rho_+g^{-1})\pi_+(g\rho_+g^{-1})^{\top}+\nonumber \right.\\
&\left.\left. +\pi_+(g\rho_+g^{-1})^{\top}\pi_+(g\rho_+g^{-1})+(\pi_-(g\rho_+g^{-1})^{\top})\right)^{\top}\delta_{\alpha}
 \right)^k.\nonumber
\end{align}
Further, using (\ref{p1}), (\ref{x1}) and
\begin{equation}
\label{a4abc}
\eta_{\alpha}\rho_0\delta_{\alpha}=\alpha_{0\infty}\rho_0,
\end{equation}
where $\rho_0\in{\cal L}^2_0$,  we find
\begin{align}
&  I^k_{\alpha}\left( \left( Ad^{+,\alpha}\right)^*_{g^{-1}}\rho_+\right)=
 Tr\left(\alpha_{0\infty}\left(\pi_+(g\rho_+g^{-1})+\pi_{-,0}(g\rho_+g^{-1})\right)^2-\right.\nonumber\\
&-
\left(\pi_+(g\rho_+g^{-1})+\pi_{-,0}(g\rho_+g^{-1})\right)
\eta_{\alpha}\left((\pi_+(g\rho_+g^{-1}))^{\top}+(\pi_{-,0}(g\rho_+g^{-1}))^{\top}\right)\delta_{\alpha}-\nonumber\\
& -
\eta_{\alpha}\left((\pi_+(g\rho_+g^{-1}))^{\top}+(\pi_{-,0}(g\rho_+g^{-1}))^{\top}\right)\delta_{\alpha}
\left(\pi_+(g\rho_+g^{-1})+\pi_{-,0}(g\rho_+g^{-1})\right)+\nonumber\\
& \left. +\eta_{\alpha}
\left(
\pi_+(g\rho_+g^{-1})+\pi_{-,0}(g\rho_+g^{-1})
\right)^{2\top}\delta_{\alpha}
 \right)^k=\nonumber\\
&=
Tr
\left(\alpha_{0\infty}(g\rho_+g^{-1})^2-g\rho_+g^{-1}\eta_{\alpha}(g\rho_+g^{-1})^{\top}\delta_{\alpha}-\right.\\
&
-\left.\eta_{\alpha}(g\rho_+g^{-1})^{\top}\delta_{\alpha}g\rho_+g^{-1}+\eta_{\alpha}\left(g\rho_+g^{-1}\right)^{2\top}\delta_{\alpha}
\right)^{k}
=I^k_{\alpha}\left(\rho_+\right).\nonumber
\end{align}
\hspace*{12cm}$\square$
\\

For $\alpha_{0\infty}\neq 0$, i.e. if $a_i\neq 0$ for all $i\in\mathbb{N}\cup\{0\}$, one has $\delta_{\alpha}=\alpha_{0\infty}\eta_{\alpha}^{-1}$ and thus
\begin{equation}
 \label{x3}
I^k_{\alpha}\left(\rho_+\right)=\alpha_{0\infty}^k
Tr
\left(\rho_+-\eta_{\alpha}\rho_+^{\top}\eta_{\alpha}^{-1}\right)^{2k}.
\end{equation}
Therefore, in nonsingular case, the Banach Lie group $G{\cal A}^2_{\alpha}$ is
isomorphic to the real orthogonal group ${\cal O}$ and the Casimirs (\ref{x3})
are obtained by this isomorphism from Casimirs of ${\cal O}$, see \cite{7}. In
singular case when $a_i=0$, $i\in I$, for some finite subset
$I\subset\mathbb{N}\cup \{0\}$ of indices,
 the expression (\ref{a30}) reduces to the following one
\begin{equation}
 \label{x4}
I^k_{\alpha}\left(\rho_+\right)=2(-1)^kTr\left(\rho_+\eta_{\alpha}\rho_+^{\top}\delta_{\alpha}\right)^k.
\end{equation}
If $a_i=0$, $i\in I$, for an infinite subset $I\subset\mathbb{N}\cup \{0\}$ of indices then
$I^k_{\alpha}(\rho_+)\equiv 0 $.


The set of invariants obtained from (\ref{x4}) is not complete for the singular
case. We can find additional invariants using methods which will be illustrated
on the $6$--dimensional case, i.e. when dim$_{\mathbb{R}}{\cal H}=6$.

Let us take the following Casimirs
\begin{align}
\label{1x}
I^1_{\alpha}(\rho_+)= & -2\left(a_0a_1a_2a_3\rho^2_{45}+a_0a_1a_2\rho^2_{35}+a_0a_1a_2a_4\rho^2_{34}+
a_0a_1\rho^2_{25}+\right. \\
 & +a_0a_1a_4\rho^2_{24}+a_0a_1a_3a_4\rho^2_{23}+a_0\rho^2_{15}+a_0a_4\rho^2_{14}+
a_0a_3a_4\rho^2_{13}+\nonumber\\
& \left. +a_0a_2a_3a_4\rho^2_{12}+\rho^2_{05}+a_4\rho^2_{04}+a_3a_4\rho^2_{03}
+a_2a_3a_4\rho^2_{02}+a_1a_2a_3a_4\rho^2_{01}\nonumber
 \right),
\end{align}
\begin{align}
\label{2x}
 C^2_{\alpha}(\rho_+):= & \frac{1}{4a_0a_4}\left(2\left(\frac{I^1_{\alpha}(\rho_+)}{2}\right)^2-I^2_{\alpha}(\rho_+)\right)=\nonumber\\
 & =a_0a^2_1a_2(a_3\rho_{23}\rho_{45}-\rho_{24}\rho_{35}+\rho_{25}\rho_{34})^2+\nonumber\\
& + a_0a_1a_2(a_3\rho_{13}\rho_{45}-\rho_{14}\rho_{35}+\rho_{15}\rho_{34})^2+\nonumber\\
& + a_0a_1(a_2a_3\rho_{12}\rho_{45}-\rho_{14}\rho_{25}+\rho_{15}\rho_{24})^2+\nonumber\\
& + a_1a_2(a_3\rho_{03}\rho_{45}-\rho_{04}\rho_{35}+\rho_{05}\rho_{34})^2+\nonumber\\
& + a_1(a_2a_3\rho_{02}\rho_{45}-\rho_{04}\rho_{25}+\rho_{05}\rho_{24})^2+\nonumber\\
& + (a_1a_2a_3\rho_{01}\rho_{45}-\rho_{04}\rho_{15}+\rho_{05}\rho_{14})^2+\nonumber\\
& + a_0a_1a_3(a_2\rho_{12}\rho_{35}-\rho_{13}\rho_{25}+\rho_{15}\rho_{23})^2+\\
& + a_1a_3(a_2\rho_{02}\rho_{35}-\rho_{03}\rho_{25}+\rho_{05}\rho_{23})^2+\nonumber\\
& + a_3(a_1a_2\rho_{01}\rho_{35}-\rho_{03}\rho_{15}+\rho_{05}\rho_{13})^2+\nonumber\\
& + a_0a_1a_3a_4(a_2\rho_{12}\rho_{34}-\rho_{13}\rho_{24}+\rho_{14}\rho_{23})^2+\nonumber\\
& + a_1a_3a_4(a_2\rho_{02}\rho_{34}-\rho_{03}\rho_{24}+\rho_{04}\rho_{23})^2+\nonumber\\
& + a_3a_4(a_1a_2\rho_{01}\rho_{34}-\rho_{03}\rho_{14}+\rho_{04}\rho_{13})^2+\nonumber\\
& + a_2a_3(a_1\rho_{01}\rho_{25}-\rho_{02}\rho_{15}+\rho_{05}\rho_{12})^2+\nonumber\\
& + a_2a_3a_4(a_1\rho_{01}\rho_{24}-\rho_{02}\rho_{14}+\rho_{04}\rho_{12})^2+\nonumber\\
& + a_2a_3^2a_4(a_1\rho_{01}\rho_{23}-\rho_{02}\rho_{13}+\rho_{03}\rho_{12})^2,\nonumber
\end{align}
\begin{align}
\label{3x}
C^3_{\alpha}(\rho_+):= & \frac{1}{a^2_0a_1a_3a^2_4}\left(2\left(\frac{I^1_{\alpha}(\rho_+)}{2}\right)^3-I^3_{\alpha}(\rho_+)-
3a_0a_1C^1_{\alpha}(\rho_+)I^1_{\alpha}(\rho_+)\right)=\nonumber\\
= & \left( a_1a_2a_3\rho_{01}\rho_{23}\rho_{45}
-a_2a_3\rho_{02}\rho_{13}\rho_{45}
+a_2a_3\rho_{03}\rho_{12}\rho_{45}-\nonumber\right.\\
- & a_1a_2\rho_{01}\rho_{24}\rho_{35}
+ a_2\rho_{02}\rho_{14}\rho_{35}
-a_2\rho_{04}\rho_{12}\rho_{35}+\nonumber\\
+ & a_1a_2\rho_{01}\rho_{25}\rho_{34}
-a_2\rho_{02}\rho_{15}\rho_{34}
+a_2\rho_{05}\rho_{12}\rho_{34}-\\
- & \rho_{03}\rho_{14}\rho_{25}
+\rho_{04}\rho_{13}\rho_{25}
+\rho_{03}\rho_{15}\rho_{24}-\nonumber\\
- & \left.\rho_{05}\rho_{13}\rho_{24}
-\rho_{04}\rho_{15}\rho_{23}
+\rho_{05}\rho_{14}\rho_{23}\right)^2.\nonumber
\end{align}
Now, assuming in (\ref{1x}), (\ref{2x}) and (\ref{3x}) that  $a_0=0$ we obtain the
following nontrivial Casimirs
\begin{align}
I^1_{\alpha}(\rho_+)= & -2\left(\rho^2_{05}+a_4\rho^2_{04}+a_3a_4\rho^2_{03}
+a_2a_3a_4\rho^2_{02}+a_1a_2a_3a_4\rho^2_{01}
 \right),
\end{align}
\begin{align}
 C^2_{\alpha}(\rho_+)=
&  a_1a_2(a_3\rho_{03}\rho_{45}-\rho_{04}\rho_{35}+\rho_{05}\rho_{34})^2+\nonumber\\
& + a_1(a_2a_3\rho_{02}\rho_{45}-\rho_{04}\rho_{25}+\rho_{05}\rho_{24})^2+\nonumber\\
& + (a_1a_2a_3\rho_{01}\rho_{45}-\rho_{04}\rho_{15}+\rho_{05}\rho_{14})^2+\nonumber\\
& + a_1a_3(a_2\rho_{02}\rho_{35}-\rho_{03}\rho_{25}+\rho_{05}\rho_{23})^2+\nonumber\\
& + a_3(a_1a_2\rho_{01}\rho_{35}-\rho_{03}\rho_{15}+\rho_{05}\rho_{13})^2+\\
& + a_1a_3a_4(a_2\rho_{02}\rho_{34}-\rho_{03}\rho_{24}+\rho_{04}\rho_{23})^2+\nonumber\\
& + a_3a_4(a_1a_2\rho_{01}\rho_{34}-\rho_{03}\rho_{14}+\rho_{04}\rho_{13})^2+\nonumber\\
& + a_2a_3(a_1\rho_{01}\rho_{25}-\rho_{02}\rho_{15}+\rho_{05}\rho_{12})^2+\nonumber\\
& + a_2a_3a_4(a_1\rho_{01}\rho_{24}-\rho_{02}\rho_{14}+\rho_{04}\rho_{12})^2+\nonumber\\
& + a_2a_3^2a_4(a_1\rho_{01}\rho_{23}-\rho_{02}\rho_{13}+\rho_{03}\rho_{12})^2,\nonumber
\end{align}
\begin{align}
C^3_{\alpha}(\rho_+)
= & \left( a_1a_2a_3\rho_{01}\rho_{23}\rho_{45}
-a_2a_3\rho_{02}\rho_{13}\rho_{45}
+a_2a_3\rho_{03}\rho_{12}\rho_{45}-\nonumber\right.\\
- & a_1a_2\rho_{01}\rho_{24}\rho_{35}
+ a_2\rho_{02}\rho_{14}\rho_{35}
-a_2\rho_{04}\rho_{12}\rho_{35}+\nonumber\\
+ & a_1a_2\rho_{01}\rho_{25}\rho_{34}
-a_2\rho_{02}\rho_{15}\rho_{34}
+a_2\rho_{05}\rho_{12}\rho_{34}-\\
- & \rho_{03}\rho_{14}\rho_{25}
+\rho_{04}\rho_{13}\rho_{25}
+\rho_{03}\rho_{15}\rho_{24}-\nonumber\\
- & \left.\rho_{05}\rho_{13}\rho_{24}
-\rho_{04}\rho_{15}\rho_{23}
+\rho_{05}\rho_{14}\rho_{23}\right)^2,\nonumber
\end{align}
for the corresponding singular case. For $a_1=0$ we find from (\ref{1x}),
(\ref{2x}) and (\ref{3x}) that
\begin{align}
I^1_{\alpha}(\rho_+)= & -2\left(a_0\rho^2_{15}+a_0a_4\rho^2_{14}+
a_0a_3a_4\rho^2_{13}+a_0a_2a_3a_4\rho^2_{12}+\right.\\
& \left.+\rho^2_{05}+a_4\rho^2_{04}+a_3a_4\rho^2_{03}
+a_2a_3a_4\rho^2_{02}\nonumber
 \right),
\end{align}
\begin{align}
 C^2_{\alpha}(\rho_+)=
& + (-\rho_{04}\rho_{15}+\rho_{05}\rho_{14})^2
 + a_3(-\rho_{03}\rho_{15}+\rho_{05}\rho_{13})^2+\nonumber\\
& + a_3a_4(-\rho_{03}\rho_{14}+\rho_{04}\rho_{13})^2
 + a_2a_3(-\rho_{02}\rho_{15}+\rho_{05}\rho_{12})^2+\\
& + a_2a_3a_4(-\rho_{02}\rho_{14}+\rho_{04}\rho_{12})^2
 + a_2a_3^2a_4(-\rho_{02}\rho_{13}+\rho_{03}\rho_{12})^2,\nonumber
\end{align}
\begin{align}
C^3_{\alpha}(\rho_+)
= & \left(
-a_2a_3\rho_{02}\rho_{13}\rho_{45}
+a_2a_3\rho_{03}\rho_{12}\rho_{45}+ a_2\rho_{02}\rho_{14}\rho_{35}-\nonumber\right.\\
& -a_2\rho_{04}\rho_{12}\rho_{35}-a_2\rho_{02}\rho_{15}\rho_{34}
+a_2\rho_{05}\rho_{12}\rho_{34}-\\
- & \rho_{03}\rho_{14}\rho_{25}
+\rho_{04}\rho_{13}\rho_{25}
+\rho_{03}\rho_{15}\rho_{24}-\nonumber\\
- & \left.\rho_{05}\rho_{13}\rho_{24}
-\rho_{04}\rho_{15}\rho_{23}
+\rho_{05}\rho_{14}\rho_{23}\right)^2,\nonumber
\end{align}
respectively. Similar analysis works in the higher dimensions. However it is
technically much more complicated.



Any two brackets of the form (\ref{p8}) always give a Poisson pencil, i.e.
\begin{equation}
 \label{p12}
c_1\{.,.\}_{\eta_{1}}+c_2\{.,.\}_{\eta_2}=\{.,.\}_{c_1\eta_1+c_2\eta_2},
\end{equation}
where $c_1, c_2\in\mathbb{R}$ and $\eta_1,\eta_2$ are some diagonal
operators on ${\cal H}$. However it is not valid in general case for
endomorphisms $\alpha :{\cal L}^{2}_+\rightarrow {\cal L}^{2}_+$  and  $\beta
:{\cal L}^{2}_+\rightarrow {\cal L}^{2}_+$ defined in  (\ref{12}). The following
proposition gives the conditions on $\alpha$ and $\beta$ which allow them to
form a pencil.

\begin{lema}
\label{a14a}
One has the following equivalent conditions:
\begin{enumerate}[(i)]
\item
\begin{equation}
\label{a14}
p\{.,.\}_{+,\alpha}+(1-p)\{.,.\}_{+,\beta}=\{.,.\}_{+,p\alpha+(1-p)\beta}
\end{equation}
for $p\in [0,1]$;
\item
\begin{equation}
\label{a15}
\left( \alpha-\beta \right)(x_+)\left( \alpha-\beta \right)(y_+)=0
\end{equation}
for $x_+,y_+\in {\cal L}_+^{2}$;
\item
\begin{equation}
\label{a16}
\left( \alpha_{ij}-\beta_{ij}\right)\left(\alpha_{jn}-\beta_{jn}\right)=0
\end{equation}
for $0\leq i<j<n$;
\item
\begin{equation}
\label{a17}
\left( a_i\ldots a_{j-1}-b_i\ldots b_{j-1}\right)\left( a_{j}-b_{j}\right)=0
\end{equation}
for  $0\leq i<j$ and $j\in\mathbb{N}\cup\{ 0\}$;
\item
\begin{equation}
\label{a18}
\left(\left(aS\right)^k-\left(bS\right)^k\right)\left( aS-bS\right)=0
\end{equation}
for $k\in\mathbb{N}$.
\end{enumerate}
\end{lema}
{\bf Proof.}
We prove the implication $(i)\Longleftrightarrow (ii) \Longleftrightarrow (iii)$  by direct calculation. Using
 the equality (\ref{11a}) and putting $j=k+i$ and $n=k+i+1$ in (\ref{a16})  we obtain
 formula (\ref{a17}),
 which implies $(iii)\Longrightarrow (iv)$. Expressing (\ref{a18}) in the basis
 $\left\{|i\rangle \langle j|\right\}_{i,j=0}^{\infty}$ we show that $(iv)\Longleftrightarrow (v)$.
 The condition (\ref{a18}) is equivalent to
 \begin{equation}
 \label{a19}
 pA+(1-p)B=\sum_{k=0}^{\infty}\left(\left(pa+(1-p)b\right)S\right)^k,
 \end{equation}
 where $A$ and $B$ are given by (\ref{11}). The condition (\ref{a19}) implies (\ref{a14}). So, we have
 $(i)\Longleftrightarrow (ii) \Longleftrightarrow (iii) \Longrightarrow (iv) \Longleftrightarrow (v) \Longrightarrow (i)$.

\hspace*{12cm}$\square$
\\


In order to solve the equations (\ref{a17}), we note that from
\begin{equation}
\left(a_i\ldots a_{j-2}-b_i \ldots b_{j-2}\right)\left(a_{j-1}-b_{j-1}\right)=0
\end{equation}
one has
\begin{equation}
\label{?}
 a_i\ldots a_{j-1}-b_i\ldots b_{j-1}=
\end{equation}
$$ =\left( a_i\ldots a_{j-2}-b_i\ldots b_{j-2}\right) b_{j-1}+
b_i\ldots b_{j-2}\left(a_{j-1}-b_{j-1}\right).
$$
Iterating (\ref{?}), we obtain
\begin{align}
\label{a22b}
a_i\ldots a_{j-1}-b_i\ldots b_{j-1} & =b_i\big( b_{i+1}\ldots b_{j-2} \left(b_{j-1}-a_{j-1}\right)+\\
& + b_{i+1}\ldots b_{j-3}\left( b_{j-2}-a_{j-2}\right)b_{j-1}+\ldots +\nonumber \\
& +
\left(a_{i+1}-b_{i+1}\right)b_{i+2}\ldots b_{j-1}\big)+\left(a_{i}-b_{i}\right)b_{i+1}\ldots b_{j-1}.\nonumber
\end{align}
Substituting (\ref{a22b}) into (\ref{a17}) one has
\begin{equation}
\label{a22b1}
\left( a_i-b_i\right)b_{i+1}\ldots b_{j-1}\left( a_j-b_j\right)=0
\end{equation}
for $j\in \mathbb{N}\cup \{ 0\}$ and $0\leq i<j$. The system of equations (\ref{a22b1}) is equivalent
to (\ref{a17}) as well as to
\begin{equation}
\label{a22b2}
\left( a_i-b_i\right)a_{i+1}\ldots a_{j-1}\left( a_j-b_j\right)=0
\end{equation}
for $j\in \mathbb{N}\cup \{ 0\}$ and $0\leq i<j$.

Now, let $I_0$ and $I$ be the subsets of $\mathbb{N}\cup \{ 0\}$ defined as
follows $I_0:=\{ i \in \mathbb{N}\cup \{ 0\} : a_i=0\;\; \textrm{or}\;\; b_i=0\}$ and
$I:=\big( \mathbb{N}\cup \{ 0\}\big)\setminus I_0$. Let us take the partition $
I=\bigcup_{l=1}^{L} I_l$
 of $I$ in the sum of the intervals $I_l:=\{ n_l+1,n_l+2,\ldots ,m_l-1\}$, where $n_l,m_l\in \mathbb{N}\cup \{ 0\}$ and
 $n_l<m_l$, for $l>1$. For $l=1$ we put $n_l=-1$. If $I_0=\emptyset$ then $L=1$ and
 $I_1=\mathbb{N}\cup \{ 0\}$.

The system (\ref{a22b1}) splits into the subsystems
\begin{equation}
\label{a22b3c}
\big(a_i-b_i\big)\big(a_j-b_j\big)=0
\end{equation}
for $i,j\in \{ n_l,n_l+1,\ldots , m_l-1, m_l\}=I_l\cup \{ n_l, m_l\}$, where $i<j$ and $l\in \{1,\ldots ,L\}$.
Let us note here that $L \in\mathbb{N}\cup \{ 0\}$.

  Summing up  we obtain:
\begin{lema}
\label{a22b3}
The sequences $\{ a_0,a_1,\ldots\}$ and $\{ b_0, b_1,\ldots \}$ give a solution of (\ref{a17}) iff for any
$l\in \{1,\ldots ,L\}$ there exists at most  one $k_l\in I_l\cup \{ n_l, m_l\}$ such that $a_{k_l}\neq b_{k_l}$.
\end{lema}

The  proposition  formulated now will be useful for the subsequent applications.

\begin{lema}
\label{as1}
Let $\{.,.\}_{+,\alpha}$ and $\{.,.\}_{+,\beta}$ form a pencil of Poisson brackets, i.e. for
sequences $\{ a_0,a_1,\ldots\}$ and $\{ b_0, b_1,\ldots \}$ and $l\in \{1,\ldots ,L\}$ there are
at most  one $k_l\in I_l\cup \{ n_l, m_l\}$ such that $a_{k_l}\neq b_{k_l}$. Then for
\begin{equation}
\label{as2}
\begin{array}{cccccc}
0\leq i<j\leq k_1   & \textrm{or}           & k_1<i<j\leq k_2  &   \textrm{or}                & k_{L-1}<i<j\leq k_L
\end{array}
\end{equation}
one has
\begin{equation}
\label{as3}
\{\rho_{ij},h\}_{+,\alpha}=\{\rho_{ij},h\}_{+,\beta},
\end{equation}
where $h\in C^{\infty}({\cal L}^2_{+})$ and $\rho_{ij}$ are coordinate functions of
$\rho_+=\sum_{0\leq i<j}\rho_{ij}|i\rangle\langle j|$.
\end{lema}

{\bf Proof.} From (\ref{a38}) we obtain
\begin{equation}
\label{a38as} \{ \rho_{ij}, h\}_{+, \alpha}-\{\rho_{ij},h\}_{+,\beta}=
\end{equation}
$$=
 \sum_{n=i+1}^{j-1}
\left( (\alpha_{nj}-\beta_{nj})
\rho_{in}\frac{\partial h}{\partial\rho_{nj}}-
(\alpha_{in}-\beta_{in})\frac{\partial h}{\partial \rho_{in}}\rho_{nj}\right)+$$ $$
+(\alpha_{ij}-\beta_{ij})\left(
\sum_{n=j+1}^{\infty} \rho_{jn}\frac{\partial h}{\partial \rho_{in}}
-\sum_{n=0}^{i-1}\frac{\partial h}{\partial \rho_{nj}}\rho_{ni}
\right)=$$
$$=
\sum_{n=i+1}^{j-1}
\left( (\alpha_{nj}-\beta_{nj})
\rho_{in}\frac{\partial h}{\partial\rho_{nj}}-
(\alpha_{in}-\beta_{in})\frac{\partial h}{\partial \rho_{in}}\rho_{nj}\right).
$$
The last equality in (\ref{a38as}) is valid since $\alpha_{ij}=\beta_{ij}$ if $i$ and
$j$ satisfy (\ref{as2}). The coefficients $\alpha_{nj}-\beta_{nj}$ and
$\alpha_{in}-\beta_{in}$  are different from zero iff $n\in\{k_1,\ldots, k_L,
k_1+1,\ldots k_L+1\}$ but it is impossible since in (\ref{a38as}) one has
$i<n<j$. \hspace*{12cm}$\square$
\\

In order to obtain the system of  integrals in involution by the Magri method
\cite{5}, let us take on ${\cal L}^2_{+}$ the Poisson bracket
\begin{equation}
\label{a23}
\{.,.\}_{+,\alpha+\epsilon\beta}=\{.,.\}_{+,\alpha}+\epsilon\{.,.\}_{+,\beta},
\end{equation}
where $\epsilon=\frac{1-p}{p}$ and $\alpha$, $\beta$ are solutions
of (\ref{a15}).

Substituting
$p\alpha+(1-p)\beta=p\left(\alpha+\epsilon\beta\right)$ into
(\ref{a30}) in place of $\alpha$
 we find Casimirs for $\left( {\cal L}_+^2,\{ .,.\}_{+,\alpha +\epsilon \beta}\right)$:
 \begin{align}
 \label{a32}
p^{2k}I^k_{\alpha+\epsilon\beta}(\rho_+)= & Tr
\left[(1+\epsilon)\left(\alpha_{0\infty}+\epsilon \beta_{0\infty}\right)\rho_+^2-\right.\\
& -\rho_+\left(\eta_{\alpha}+\epsilon\eta_{\beta}\right)\rho_+^{\top}
\left(\delta_{\alpha}+\epsilon\delta_{\beta}\right)-\nonumber\\
&-\left(\eta_{\alpha}+\epsilon\eta_{\beta}\right)\rho_+^{\top}\left(\delta_{\alpha}+\epsilon\delta_{\beta}\right)\rho_+ +
\nonumber\\
&\left.
+\left(\eta_{\alpha}+\epsilon\eta_{\beta}\right)\left(\rho_+^{\top}\right)^2\left(\delta_{\alpha}+\epsilon\delta_{\beta}\right)
\right]^{k}.\nonumber
 \end{align}
Expanding $I^k_{\alpha+\epsilon\beta}(\rho_+)$  whith respect to
the parameter $\epsilon$
\begin{equation}
\label{a24}
I_{\alpha+\epsilon\beta}^k(\rho_+)=\sum_{n=0}^{2k}h^k_{n}(\rho_+)\epsilon^n
\end{equation}
we obtain from  (\ref{a23}) the system of integrals $h^k_{n}$ wich are in involution
\begin{align}
\label{a27} & \{ h^k_{n},h^l_{m}\}_{+,\alpha}=0= \{
h^k_{n},h^l_{m}\}_{+,\beta},
\end{align}
where $k,l,n,m\in\mathbb{N}\cup \{0\}$ and $ n\leq 2k, m\leq 2l$.

In such a way, using (\ref{a38}) we obtain  a  hierarchy of Hamilton equations
\begin{align}
\label{a39} \frac{d \rho_{ij}}{d t}= &
 \sum_{n=i+1}^{j-1}
\left( \alpha_{nj}\rho_{in}\frac{\partial h^k_{m}}{\partial
\rho_{nj}}-
\alpha_{in}\frac{\partial h^k_{m}}{\partial \rho_{in}}\rho_{nj}\right)+\\
& +\sum_{n=0}^{i-1} \left( \rho_{nj}\frac{\partial h^k_{m}}{\partial
\rho_{ni}}-
\alpha_{ij}\frac{\partial h^k_{m}}{\partial \rho_{nj}}\rho_{ni}\right)+\nonumber\\
& +\sum_{n=j+1}^{\infty} \left( \alpha_{ij}\rho_{jn}\frac{\partial
h^k_{m}}{\partial \rho_{in}}- \frac{\partial h^k_{m}}{\partial
\rho_{jn}}\rho_{in}\right),\nonumber
\end{align}
indexed by $k\in\mathbb{N}$ and $m\leq 2k$.

\begin{lema}
\label{a40}
Let $i,j\in\mathbb{N}\cup \{ 0\}$ satisfy  condition (\ref{as2}). Then $\rho_{ij}$ are constants of motion
\begin{equation}
\label{a41}
\frac{d \rho_{ij}}{d t}=\{ \rho_{ij}, h^k_m\}_{+,\alpha}=0
\end{equation}
for any Hamiltonian $h^k_m$ given by (\ref{a24}).
\end{lema}

{\bf Proof.} It follows from (\ref{a24}) that $\{ \rho_{ij}, h^k_m\}_{+,\alpha}=0$ iff
\begin{equation}
\label{a42}
\{ \rho_{ij}, I^k_{\alpha+\epsilon\beta}\}_{+,\alpha}=0.
\end{equation}
In order to verify (\ref{a42}) we note that
\begin{equation}
\label{a43}
0=\{ \rho_{ij}, I^k_{\alpha+\epsilon\beta}\}_{+,\alpha+\epsilon\beta}=
\{ \rho_{ij}, I^k_{\alpha+\epsilon\beta}\}_{+,\alpha}+\epsilon
\{ \rho_{ij}, I^k_{\alpha+\epsilon\beta}\}_{+,\beta}.
\end{equation}
From  Proposition \ref{as1} we have
\begin{equation}
\label{a44}
\{ \rho_{ij}, I^k_{\alpha+\epsilon\beta}\}_{+,\alpha}=
\{ \rho_{ij}, I^k_{\alpha+\epsilon\beta}\}_{+,\beta}
\end{equation}
iff $i,j\in\mathbb{N}\cup \{ 0\}$ satisfy (\ref{as2}). Now (\ref{a43}) and (\ref{a44}) imply (\ref{a42}).

\hspace*{12cm}$\square$

We shall apply the above propositions to the examples investigated in the next
section.

\section{Examples of Hamilton equations and their solutions}

Let us now illustrate the Hamiltonian  hierarchy (\ref{a39}) by some examples.
We shall consider the case $a_i=1$ for $i\in \mathbb{N}\cup \{ 0\}$, $b_1=b$
and $b_i=1$ for $i\neq 1$. It is convenient to use the block matrix notation for this
case, i.e.
\begin{equation}
\label{a45}
\rho_+=\left(\begin{array}{cc|c}
0 & a & x^{\top}\\
0 & 0 & y^{\top}\\
\hline
{\bf 0} & {\bf 0} & {\boldsymbol \delta}
\end{array}\right),
\end{equation}
where $a\in\mathbb{R}$, $x,y\in  l^{2}$ and
\begin{equation}
\label{a46}
{\boldsymbol \delta}=\left(\begin{array}{cccc}
0 & {\boldsymbol \delta}_{12} & {\boldsymbol \delta}_{13} & \ldots \\
0 & 0           & {\boldsymbol \delta}_{23} & \ldots \\
0 & 0           & 0           & \ldots \\
\vdots &\vdots    & \vdots       & \ddots
\end{array}\right)
\in {\cal L}^2_+\left( l^2\right),
\end{equation}
where $l^2$ is the Hilbert space of square summable real sequences. The
Poisson bracket (\ref{29})  for $f,g,h\in C^{\infty}({\cal L}^2_+)$ and the
Hamilton equations (\ref{a39}) in the matrix coordinates $(a,x,y,{\boldsymbol
\delta} )$ take the following form
\begin{equation}
\label{a47}
\{ f,g\}_{+,\alpha}= a\left( \frac{\partial g}{\partial y^{\top}}\frac{\partial f}{\partial x}-
\frac{\partial f}{\partial y^{\top}}\frac{\partial g}{\partial x}\right)+
\end{equation}
$$
+x^{\top}\left(\left(\left( \frac{\partial f}{\partial {\boldsymbol \delta}}\right)^{\top}-\frac{\partial f}{\partial {\boldsymbol \delta}}\right)\frac{\partial g}{\partial x}-
\left(\left( \frac{\partial g}{\partial {\boldsymbol \delta}}\right)^{\top}-\frac{\partial g}{\partial {\boldsymbol \delta}}\right)\frac{\partial f}{\partial x}+
\frac{\partial f}{\partial y}\frac{\partial g}{\partial a}-\frac{\partial g}{\partial y}\frac{\partial f}{\partial a}\right)+
$$$$
+ y^{\top}\left(\left(\left( \frac{\partial f}{\partial {\boldsymbol \delta}}\right)^{\top}-\frac{\partial f}{\partial {\boldsymbol \delta}}\right)\frac{\partial g}{\partial y}-
\left(\left( \frac{\partial g}{\partial {\boldsymbol \delta}}\right)^{\top}-\frac{\partial g}{\partial {\boldsymbol \delta}}\right)\frac{\partial f}{\partial y}-
\frac{\partial f}{\partial x}\frac{\partial g}{\partial a}+\frac{\partial g}{\partial x}\frac{\partial f}{\partial a}\right)+
$$$$
+Tr \left( {\boldsymbol \delta}\left( -\left( \frac{\partial f}{\partial x}\frac{\partial g}{\partial x^{\top}}-
\frac{\partial g}{\partial x}\frac{\partial f}{\partial x^{\top}}\right)-
\left( \frac{\partial f}{\partial y}\frac{\partial g}{\partial y^{\top}}-
\frac{\partial g}{\partial y}\frac{\partial f}{\partial y^{\top}}\right)+\right.\right.
$$$$
+\left(\left( \frac{\partial f}{\partial {\boldsymbol \delta}}\right)^{\top}-\frac{\partial f}{\partial {\boldsymbol \delta}}\right)
\left(\left( \frac{\partial g}{\partial {\boldsymbol \delta}}\right)^{\top}-\frac{\partial g}{\partial {\boldsymbol \delta}}\right)-
$$$$
\left.\left.
-\left(\left( \frac{\partial g}{\partial {\boldsymbol \delta}}\right)^{\top}-\frac{\partial g}{\partial {\boldsymbol \delta}}\right)
\left(\left( \frac{\partial f}{\partial {\boldsymbol \delta}}\right)^{\top}-\frac{\partial f}{\partial {\boldsymbol \delta}}\right) \right)\right)
$$
and
\begin{align}
\label{a48}
& \frac{da}{dt}=y^{\top}\frac{\partial h}{\partial x} -x^{\top}\frac{\partial h}{\partial y},\\
\label{a48c1}& \frac{dx}{dt}=-\frac{\partial h}{\partial a}y+({\boldsymbol \delta}-{\boldsymbol \delta}^{\top})\frac{\partial h}{\partial x}+a\frac{\partial h}{\partial y}-\left( \frac{\partial h}{\partial {\boldsymbol \delta}}-\frac{\partial h}{\partial {\boldsymbol \delta}^{\top}}\right) x,\\
\label{a48c2} & \frac{dy}{dt}=\frac{\partial h}{\partial a}x+({\boldsymbol \delta}-{\boldsymbol \delta}^{\top})\frac{\partial h}{\partial x}-a\frac{\partial h}{\partial y}-\left( \frac{\partial h}{\partial {\boldsymbol \delta}}-\frac{\partial h}{\partial {\boldsymbol \delta}^{\top}}\right) y,\\
\label{a48c3} & \frac{d{\boldsymbol \delta}}{dt}=\pi_{+,\alpha}\left(
\frac{\partial h}{\partial x}x^{\top}+\frac{\partial h}{\partial y}y^{\top}+\left[{\boldsymbol \delta},\frac{\partial h}{\partial {\boldsymbol \delta}}-\frac{\partial h}{\partial {\boldsymbol \delta}^{\top}}\right]
\right),
\end{align}
respectively. If the Hamiltonian $h\in C^{\infty}({\cal L}^2_+)$ in (\ref{a48}-\ref{a48c3}) is functionally
 dependent on the integrals of motion $h^k_n$, where $k\in\mathbb{N}$ and $n\leq 2k$, we have
\begin{equation}
\label{a49}
 \frac{da}{dt}=0\;\;\;\;\; \textrm{and}\;\;\;\;\; \frac{d{\boldsymbol \delta}}{dt}=0,
\end{equation}
from Proposition (7).
So, in this case the system of Hamilton equations (\ref{a48}-\ref{a48c3}) reduces to equations (\ref{a48c1}-\ref{a48c2}) on the vector valued functions $x(t)$ and $y(t)$ in which, due to (\ref{a49}), the quantities $a$ and ${\boldsymbol \delta}$ play a role of the parameters constant in $t$.

We can simplify equations (\ref{a48c1}-\ref{a48c2}) by passing  to the complex vector variable $z=x+iy\in (l^2)^\mathbb{C}$. After carrying out easy calculations we find that they are equivalent to
\begin{equation}
 \label{1d}
\frac{dz}{dt}=\left(\left(\frac{\partial h}{\partial {\boldsymbol \delta}}\right)^{\top}-\frac{\partial h}{\partial {\boldsymbol \delta}}+i\frac{\partial h}{\partial a}{\mathbbm 1}\right)z+ 2\left({\boldsymbol \delta}-{\boldsymbol \delta}^{\top}-ia{\mathbbm 1}\right)\frac{\partial h}{\partial \bar{z}},
\end{equation}
where ${\mathbbm 1}$ is  identity operator.

In what follows we are looking for the solutions of (\ref{1d}) with a Hamiltonian
$h$ which will be functionally expressed by the integrals of motion obtained
from (\ref{a24}) for $k\leq2$. These integrals of motion are given by
\begin{align}
 \label{1?}
& h^1_0=h_1-2h_2,\nonumber\\
& h^1_1=(1+b)h_1-4h_2,\nonumber\\
& h^1_2=bh_1-2h_2,\nonumber\\
& h^2_0=h_3+h_4+h_5,\nonumber\\
& h^2_1=2(1+b)h_3+4h_4+(3+b)h_5,\\
& h^2_2=(1+4b+b^2)h_3+6h_4+3(1+b)h_5,\nonumber\\
& h^2_3=2b(1+b)h_3+4h_4+(1+3b)h_5,\nonumber\\
& h^2_4=b^2h_3+h_4+bh_5,\nonumber
\end{align}
where one defines the functions $h_1$, $h_2$, $h_3$, $h_4$ and $h_5$ in the following way
\begin{align}
\label{a50}
& h_1:=-2a^2+Tr({\boldsymbol \delta}-{\boldsymbol \delta}^{\top})^2,\nonumber\\
& h_2:=\bar{z}^{\top}z,\nonumber\\
& h_3:=Tr({\boldsymbol \delta}-{\boldsymbol \delta}^{\top})^4+2a^4,\\
& h_4:=|z^{\top}z|^2+(\bar{z}^{\top}z)^2,\nonumber \\
& h_5:=4a^2\bar{z}^{\top}z- 4ia\bar{z}^{\top}({\boldsymbol \delta}-{\boldsymbol \delta}^{\top})z-4\bar{z}^{\top}({\boldsymbol \delta}-{\boldsymbol \delta}^{\top})^2z.\nonumber
\end{align}
Since, functions $h_m$ can be expressed as a linear combinations of
functions $h_m^k$ they are also the integrals of motion. Note that $h_m$, for
$m=1,2,3,4,5$, are functionally independent. Note also that $h_0^1$ and
$h_0^2$ are Casimirs and they can not be considered as generators of the
nontrivial evolution.

In what follows we are looking for  solutions of (\ref{1d}) with the Hamiltonian
\begin{align}
\label{2d}
h:=  \frac 12\left(h_1-h_4\right) +(h_2)^2=
\end{align}
$$
= \frac{1}{2}\left((\bar{z}^{\top}z)^2-|z^{\top}z|^2+Tr({\boldsymbol \delta}-{\boldsymbol \delta}^{\top})^2-2a^2\right).
$$
After substitution (\ref{2d}) into (\ref{1d}) we obtain
\begin{align}
\label{3d}
\frac 12\frac{dz}{dt}= &
(1+\bar{z}^{\top}z)\left({\boldsymbol \delta}-{\boldsymbol \delta}^{\top}\right) z-ia\left( 1+\bar{z}^{\top}z\right)z-z^{\top}z({\boldsymbol \delta}-{\boldsymbol \delta}^{\top}-ia{\mathbbm 1})\bar{z},
\end{align}
where
\begin{align}
\label{4d} & \bar{z}^{\top}(t)z(t)=h_2=const=:c^2,\\
\label{5d} & |z(t)^{\top}z(t)|^2=h_4-(h_2)^2=const=:\varrho^2,\\
\label{5dd} & \bar{z}^{\top}(t)({\boldsymbol \delta}-{\boldsymbol \delta}^{\top})^2z(t)+ia\bar{z}^{\top}(t)({\boldsymbol \delta}-{\boldsymbol \delta}^{\top})z(t)=a^2h_2-\frac 14h_5=const
\end{align}
do not depend on the parameter $t$. Because of  (\ref{4d}), we see that the first two terms in right-hand side of (\ref{3d}) are linear in $z$. The integrals (\ref{5d}) and (\ref{5dd}), as we shall see later, also are useful for the further simplification of (\ref{3d}).

For the case when dim$(l^2)^{\mathbb{C}}\!=\!\infty$ it is interesting to realize
$(l^2)^{\mathbb{C}}$ as the Hilbert space ${ L}^2(\mathbb{R}, d\mu)$ of
complex valued  functions $\psi\in L^2(\mathbb{R}, d\mu)$ square-integrable
with respect to some measure $d\mu$, such one that $L^2(\mathbb{R}, d\mu )$
has the orthogonal basis consisting of the real valued functions. Then the
Hilbert--Schmidt operator $\delta-\delta^{\top}$ will be integral operator given
by some kernel $\Delta (x,y)$ such that
$\int_{\mathbb{R}\times\mathbb{R}}|\Delta(x,y)|^2dxdy<+\infty$, and the right
side of (\ref{3d}) assumes the form of nonlinear integral operator and we obtain
\begin{equation}
\label{e20eee}
 \frac 12\frac{d}{dt}\psi(x,t)=\left(1+\int_{\mathbb{R}}|\psi(y,t)|^2dy\right)\int_{\mathbb{R}}\Delta(x,y)\psi(y,t)dy-
\end{equation}
$$
-\left(\int_{\mathbb{R}}\psi^2(y,t)dy\right)\int_{\mathbb{R}}\Delta(x,y)\overline{\psi(y,t)}dt-
$$$$
-ia\left(\left(1+\int_{\mathbb{R}}|\psi(y,t)|^2dy\right)\psi(x,t)-\left(\int_{\mathbb{R}}\psi^2(y,t)dy\right)\overline{\psi(x,t)}\right).
$$
Note that the integrals of motion (\ref{a50}) one can also rewrite in the integral form.

Now, let us consider in details the finite dimensional case, i.e. $\left(l
\right)^{2\mathbb{C}} \cong\mathbb{C}^{N}$ when $N<\infty$. Passing to new
coordinates
\begin{equation}
\label{6d} \mathbb{C}^{N}\ni z\longmapsto Oz\in \mathbb{C}^{N}
\end{equation}
by the real orthogonal mapping,  i.e. $O\in Mat_{N\times N}(\mathbb{R})$ and
$OO^{\top}=O^{\top}O={\mathbbm 1}$ one can transform ${\boldsymbol
\delta}-{\boldsymbol \delta}^{\top}$ to the following $2\times 2$--blocks
matrices:
\begin{equation}
\label{7d} {\boldsymbol \delta}-{\boldsymbol \delta}^{\top}\longmapsto O\left({\boldsymbol \delta}-{\boldsymbol \delta}^{\top}\right)O^{\top}=\left(\begin{array}{ccc}
\lambda_{1}{\boldsymbol \varepsilon} & \ldots & 0\\
\vdots & \ddots & \vdots\\
0 & \ldots & \lambda_{N}{\boldsymbol \varepsilon}
\end{array}\right),
\end{equation}
 for $N=2M$, and
\begin{equation}
\label{7dd} {\boldsymbol \delta}-{\boldsymbol \delta}^{\top}\longmapsto O\left({\boldsymbol \delta}-{\boldsymbol \delta}^{\top}\right)O^{\top}=\left(\begin{array}{cccc}
\lambda_{1}{\boldsymbol \varepsilon} & \ldots & 0 & 0\\
\vdots & \ddots & \vdots & \vdots\\
0 & \ldots & \lambda_{N}{\boldsymbol \varepsilon} & 0\\
0 &\ldots & 0 & 0
\end{array}\right),
\end{equation}
for $N=2M+1$, where ${\bf {\boldsymbol \varepsilon}}=\left(\begin{array}{cc}
                          0 & 1\\
                         -1 & 0
                         \end{array}\right)$
and $\lambda_{k}\in\mathbb{R}$, see \cite{3}. In (\ref{7d}-\ref{7dd}) we allow
some $\lambda_k$ to be equal to zero.

In such a way, putting
\begin{equation}
\label{8d} Oz=:\left(\begin{array}{c}\xi_1\\\vdots\\\xi_{M}\end{array}\right)\in\mathbb{C}^{2M}\;\;\;\textrm{or}\;\;\;
Oz=:\left(\begin{array}{c}\xi_1\\\vdots\\\xi_{M}\\\xi_0\end{array}\right)\in\mathbb{C}^{2M+1},
\end{equation}
where $\xi_k\in\mathbb{C}^2$, $k=1,\ldots, M$, and $\xi\in\mathbb{C}$ we transform equation (\ref{3d}) to
\begin{align}
\label{9d} \frac 12\frac{d\xi_k}{dt}= & \lambda_k(1+c^2){\bf {\boldsymbol \varepsilon}}\xi_k-ia(1+c^2)\xi_k-\\
& -\left(\sum_{l=1}^{M}\xi_l^{\top}\xi_l\right)\left(\lambda_k{\bf {\boldsymbol \varepsilon}}-ia\mathbbm{1}\right)\bar{\xi}_k,\nonumber
\end{align}
in the even dimensional case, and to
\begin{align}
\label{10d} \frac 12\frac{d\xi_k}{dt}= & \lambda_k\left(1+c^2\right){\bf {\boldsymbol \varepsilon}}\xi_k
 -ia\left(1+c^2\right)\xi_k-\\
& -\left(\sum_{l=1}^{M}\xi_l^{\top}\xi_l+\xi_0^2\right)\left(\lambda_k{\bf {\boldsymbol \varepsilon}}-ia\mathbbm{1}\right)\bar{\xi}_k,\nonumber
\end{align}
\begin{align}
\label{11d}
\frac 12\frac{d\xi_0}{dt}= &  -ia\left(1+c^2\right)\xi_0
+ia\left(\xi_0^2+\sum_{l=1}^{M}\xi_l^{\top}\xi_l\right)\bar{\xi}_0,
\end{align}
for the odd dimensional case, where $k=1,\dots, M$

For both cases we obtain from (\ref{9d}) and (\ref{10d}) additional invariants
\begin{align}
 &  \label{12d}\lambda_k\bar{\xi}_k^{\;\top}(t) \xi_k(t)-ia\bar{\xi}_k^{\;\top}(t){\bf {\boldsymbol \varepsilon}}\xi_k(t)=:c_k=const,\\
 & \label{13d} \frac{a^2}{2}|\xi_k^{\top}(t)\xi_k(t)|^2-\frac{1}{2}(a^2-\lambda_k^2)(\bar{\xi}^{\;\top}_k(t)\xi_k(t))^2-
\lambda_kc_k\bar{\xi}^{\;\top}_k(t)\xi_k(t)=:d_k=const.
\end{align}

Since  for $\xi\in\mathbb{C}^2$ one has the identity
\begin{equation}
 \label{14d}
|\xi^{\top}_{k}\xi_k|^2=(\xi^{\top}_{k}\xi_k)^2+(\xi^{\top}_{k}{\boldsymbol \varepsilon}\xi_k)^2
\end{equation}
we find that
\begin{equation}
 \label{15d}
d_k=-\frac 12c_k^2,
\end{equation}
i.e. the above invariants are functionally dependent.

For the subsequent simplification we rewrite (\ref{9d}), (\ref{10d}) and
(\ref{11d}) in  new coordinates. These new coordinates will consist of the
complex variables
\begin{equation}
 \label{16d}
\eta_k:=\xi_k^{\top}\xi_k, \;\;\;\;\;\eta_0:=\xi_0^2
\end{equation}
as well as of the real variables $\alpha_k,\beta_k\in\mathbb{R}$. The real ones are defined by the polar decompositions
\begin{equation}
 \label{17d}
Re\; \xi_k=||Re\;\xi_k||\left(\begin{array}{c}
                               \cos \alpha_k\\
                               \sin \alpha_k
                              \end{array}\right),\;\;\;\;\;
Im\; \xi_k=||Im\;\xi_k||\left(\begin{array}{c}
                               \cos \beta_k\\
                               \sin \beta_k
                              \end{array}\right),
\end{equation}
of $Re\;\xi_k,\; Im\; \xi_k\in\mathbb{R}^2$, where
\begin{align}
 \label{18d}
& ||Re\;\xi_k||^2=\frac 12\left( Re\;\eta_k+\frac{-\lambda_kc_k\pm a\sqrt{(a^2-\lambda_k^2)|\eta_k|^2+c_k}}{a^2-\lambda_k^2}\right),\\
\label{18dd} & ||Im\;\xi_k||^2=\frac 12\left( -Re\;\eta_k+\frac{-\lambda_kc_k\pm
a\sqrt{(a^2-\lambda_k^2)|\eta_k|^2+c_k}}{a^2-\lambda_k^2}\right),
\end{align}
and $k=1\ldots ,M$. Now, after using  invariants (\ref{4d}), (\ref{5d}), (\ref{12d})
and (\ref{13d}) we transform (\ref{9d}) to the following equivalent system of
equations
\begin{align}
& \label{19d} \frac{d\eta_k}{dt}=&&-4ia(1+c^2)\eta_k
\pm 4i\sqrt{(a^2-\lambda_k^2)|\eta_k|^2+c_k^2}\sum_{l=1}^{M}\eta_l ,\\
& \label{20d} \frac{d\alpha_k}{dt}= && 2(a^2-\lambda_k^2)
\frac{\left(c_k\sum_{l=1}^M(\eta_l+\bar{\eta}_l)+\lambda_k\left( \bar{\eta}_k\sum_{l=1}^M\eta_l+\eta_k\sum_{l=1}^M\bar{\eta}_l\right)\right)}
{(a^2-\lambda_k^2)(\eta_k+\bar{\eta}_k)-2\lambda_kc_k\pm2a\sqrt{(a^2-\lambda_k^2)|\eta_k|^2+c_k^2}}-\nonumber\\
& &&
-2\lambda_k(1+c^2)
,\\
& \label{21d} \frac{d\beta_k}{dt}= && 2(a^2-\lambda_k^2)
\frac{\left(c_k\sum_{l=1}^M(\eta_l+\bar{\eta}_l)-\lambda_k\left( \bar{\eta}_k\sum_{l=1}^M\eta_l+\eta_k\sum_{l=1}^M\bar{\eta}_l\right)\right)}
{(a^2-\lambda_k^2)(\eta_k+\bar{\eta}_k)+2\lambda_kc_k\mp2a\sqrt{(a^2-\lambda_k^2)|\eta_k|^2+c_k^2}}-\nonumber\\
& &&
-2\lambda_k(1+c^2).
\end{align}
In the odd dimensional case equations (\ref{10d}) and (\ref{11d}) in the coordinates $\eta_k,\alpha_k,\beta_k$ and $\eta_0$ have the following form
\begin{align}
& \label{22d} \frac{d\eta_k}{dt}=&&-4ia(1+c^2)\eta_k
\pm 2i\sqrt{(a^2-\lambda_k^2)|\eta_k|^2+c_k^2}\sum_{l=0}^{M}\eta_l,\\
& \label{23d} \frac{d\alpha_k}{dt}= && 2(a^2-\lambda_k^2)
\frac{\left(c_k\sum_{l=0}^M(\eta_l+\bar{\eta}_l)+\lambda_k\left( \bar{\eta}_k\sum_{l=0}^M\eta_l+\eta_k\sum_{l=0}^M\bar{\eta}_l\right)\right)}
{(a^2-\lambda_k^2)(\eta_k+\bar{\eta}_k)-2\lambda_kc_k\pm2a\sqrt{(a^2-\lambda_k^2)|\eta_k|^2+c_k^2}}-\nonumber\\
& &&
-2\lambda_k(1+c^2)
,\\
& \label{24d} \frac{d\beta_k}{dt}= && 2(a^2-\lambda_k^2)
\frac{\left(c_k\sum_{l=0}^M(\eta_l+\bar{\eta}_l)-\lambda_k\left( \bar{\eta}_k\sum_{l=0}^M\eta_l+\eta_k\sum_{l=0}^M\bar{\eta}_l\right)\right)}
{(a^2-\lambda_k^2)(\eta_k+\bar{\eta}_k)+2\lambda_kc_k\mp2a\sqrt{(a^2-\lambda_k^2)|\eta_k|^2+c_k^2}}-\nonumber\\
& &&
-2\lambda_k(1+c^2)
,\\
& \label{25d}\frac{d\eta_0}{dt}=&& -4ia(1+c^2)\eta_0+4ia|\eta_0|\sum_{l=0}^M\eta_l.
\end{align}

We see from (\ref{20d}-\ref{21d}) and from (\ref{23d}-\ref{24d}) that $\frac{d}{dt}\alpha_k(t)$ and
$\frac{d}{dt}\beta_k(t)$
 are expressed by the functions $\eta_k(t)$, $k=0,1,\ldots , M$.
 Thus we can reduce the problem of solving (\ref{19d}-\ref{21d}) to solving of (\ref{19d}).
 Similarly, we reduce the problem of solving (\ref{22d}-\ref{25d}) to solving of (\ref{22d}) and (\ref{25d}).

For further simplification we replace the functions $\eta_k(t)$ by
\begin{equation}
 \label{26d}
q_k(t)+ip_k(t):=e^{-i\varphi(t)}\eta_k(t),
\end{equation}
where the real valued function $\varphi(t)$ is defined by
\begin{equation}
 \label{27d}
\varrho e^{i\varphi(t)}:=\sum_{l=1}^M\eta_l(t)
\end{equation}
for the even dimensional case and by
\begin{equation}
 \label{28d}
\varrho e^{i\varphi(t)}:=\sum_{l=0}^M\eta_l(t)
\end{equation}
for the odd dimensional case. Note that the quantity $\varrho\in\mathbb{R}$ does not depend on $t$, see (\ref{5d}). Note also that for the even  dimensional case $k=1,\ldots, M$ and for the odd dimansional case $k=0,1,\ldots, M$.

Substituting (\ref{26d}) into (\ref{19d}), we transform these to equations
\begin{align}
 \label{29d} &\frac{dq_k}{dt}=4p_k\sum_{l=1}^Mr_l,\\
\label{30d} &\frac{dp_k}{dt}=-4q_k\sum_{l=1}^Mr_l+4\varrho r_k,\\
\label{33d} &\frac{dr_k}{dt}=4\varrho(a^2-\lambda_k^2)p_k.
\end{align}
where
\begin{equation}
 \label{31d}
r_k:=\pm\sqrt{(a^2-\lambda_k^2)(q_k^2+p_k^2)+c_k^2},
\end{equation}
and $k=1,\ldots, M$. Additionally we have
\begin{align}
 \label{32d} &\frac{d\varphi}{dt}=-4a^2(1+c^2)+4\sum_{l=1}^Mr_l.
\end{align}

Further, considering $r_k(t)$ as  independent functions, instead of
(\ref{29d}-\ref{30d}) we shall investigate (\ref{29d}-\ref{33d}) as equations for
the functions $q_k(t)$, $p_k(t)$ and $r_k(t)$. Let us note that in such a case
one can consider (\ref{31d}) as an invariant for the  system (\ref{29d}-\ref{33d}).

It is easy to show that
\begin{align}
\label{35d} &\sum_{l=1}^Mp_l(t)=const=0,\\
 \label{34d} &\sum_{l=1}^Mq_l(t)=const=\varrho,\\
\label{36d} &\sum_{l=1}^M  (a^2-\lambda_l^2)\varrho q_l(t)-\frac 12\left(\sum_{l=1}^Mr_l(t)\right)^2=:g=const,\\
\label{37d} &\sum_{l=1}^M\frac{r_l(t)}{a^2-\lambda_l^2}=:f=const
\end{align}
are  invariants for this system in the even dimensional case.

For the odd dimensional case there exist similar equations and invariants. They are given also by the formulas (\ref{29d}-\ref{33d}), (\ref{32d}) and  (\ref{35d}-\ref{37d}) but with the index of sumation taken from $0$ to $M$ and $\lambda_0=0$.

The equation (\ref{25d}) is equivalent to the system of three equations
\begin{align}
 \label{38d} &\frac{dq_0}{dt}=4p_0\sum_{l=0}^Mr_l,\nonumber\\
&\frac{dp_0}{dt}=-4q_0\sum_{l=0}^Mr_l+4\varrho r_0,\\
&\frac{dr_0}{dt}=4\varrho a^2p_0.\nonumber
\end{align}
Recall here that $r_0=a|\eta_0|$ and $\lambda_0=0$. Finally we note that in
variables $q_k$, $p_k$, $r_k$ the equations in question for the both cases
have the same form and the structure of  non--linearity is more simple.

In the next part of this section we find solutions of equations (\ref{3d}) in the
dimensions $N=2,3$ and $N=4$. We shall do that by two steps. At first step we
integrate the system (\ref{29d}-\ref{33d}). Secondly, given
$\eta_k(t)=q_k(t)+ip_k(t)$ and $r_k(t)$ we obtain $\alpha_k(t)$ and
$\beta_k(t)$ from (\ref{20d}-\ref{21d}) or (\ref{23d}-\ref{24d}). Hence, by
(\ref{18d}-\ref{18dd}) we find solutions of (\ref{9d}-\ref{11d}) and finally after the
orthogonal transformation the solutions of (\ref{3d}).

\subsubsection*{ 1. Case $N=2$}

In this case equations (\ref{29d}-\ref{33d}) are reduced to the system
\begin{align}
 \label{41d} &\frac{dq_1}{dt}=4r_1p_1,\nonumber\\
&\frac{dp_1}{dt}=-4r_1q_1+4\varrho r_1,\\
&\frac{dr_1}{dt}=4\varrho (a^2-\lambda_1^2)p_1,\nonumber
\end{align}
which is solved by $q_1(t)=\varrho$, $p_1(t)=0$ and $r_1(t)=\pm\sqrt{(a^2-\lambda_1)^2\varrho^2+c_1^2}$. Since $r_1(t)=const$ we find
\begin{equation}
 \label{44d}
\varphi(t)=\omega_1t+\varphi_0
\end{equation}
and
\begin{equation}
 \label{45d}
\eta_1(t)=\varrho e^{i\varphi(t)}=\varrho e^{i(\omega_1+\varphi_0)},
\end{equation}
where
\begin{equation}
 \label{46d}
\omega_1:=-4a^2(1+c^2)+4r_1.
\end{equation}
Now, substituting (\ref{45d}) into (\ref{23d}-\ref{24d}) we get
\begin{align}
 \label{47d} & \frac{d\alpha_1}{dt}=-2\lambda_1(1+c^2)+2c_1+\frac{2\lambda_1\varrho-2\frac{c_1c^2}{\varrho}}
{\cos (\omega_1t+\varphi_0)+\frac{c^2}{\varrho}},\\
\label{48d} & \frac{d\beta_1}{dt}=-2\lambda_1(1+c^2)+2c_1-\frac{2\lambda_1\varrho+2\frac{c_1c^2}{\varrho}}
{\cos (\omega_1t+\varphi_0)-\frac{c^2}{\varrho}}.
\end{align}
Solutions of (\ref{47d}-\ref{48d}) are given by the trigonometric functions
\begin{align}
 \label{49d} & \alpha_1(t)=2(c_1-\lambda_1(1+c^2))t+
4\frac{\lambda_1\varrho-\frac{c_1c^2}{\varrho}}{\omega_1\sqrt{\frac{c^4}{\varrho^2}-1}}
arctg \frac{\left(\frac{c^2}{\varrho}-1\right) tg\frac {\omega_1t+\varphi_0}{2}}{\sqrt{\frac{c^4}{\varrho^2}-1}},\\
\label{50d}  & \beta_1(t)=2(c_1-\lambda_1(1+c^2))t-
4\frac{\lambda_1\varrho+\frac{c_1c^2}{\varrho}}{\omega_1\sqrt{\frac{c^4}{\varrho^2}-1}}
arctg \frac{\left(-\frac{c^2}{\varrho}-1\right) tg\frac {\omega_1t+\varphi_0}{2}}{\sqrt{\frac{c^4}{\varrho^2}-1}}
.
\end{align}
Finally we find from (\ref{18d}) that
\begin{align}
 \label{51d}
z(t)=& x(t)+iy(t)=\xi_1(t)=  \frac{1}{\sqrt{2}}\sqrt{\varrho\cos (\omega_1t+\varphi_0)+c^2}
\left(\begin{array}{c}
       \cos\alpha_1(t)\\
       \sin \alpha_1(t)
      \end{array}\right)+\nonumber\\
& + \frac{i}{\sqrt{2}}\sqrt{-\varrho\cos (\omega_1t+\varphi_0)+c^2}
\left(\begin{array}{c}
       \cos\beta_1(t)\\
       \sin \beta_1(t)
      \end{array}\right),
\end{align}
where $\alpha_1(t)$ and $\beta_1(t)$ are given by (\ref{49d}-\ref{50d}). Note
that $c^2\geq \varrho$, what follows from the identity (\ref{14d}). So, for the
dimension $N=2$ the equations (\ref{3d}) describe the evolution of two vectors
$x(t)$ and $y(t)$ in $\mathbb{R}^2$ having invariants
\begin{equation}
 \label{52d}
||x(t)||^2+||y(t)||^2=c^2
\end{equation}
and
\begin{equation}
 \label{53d}
||x(t)||\;||y(t)||\;\sin(\beta_1(t)-\alpha_1(t))=\pm\frac 12\sqrt{c^4-\varrho^2},
\end{equation}
 where  (\ref{53d}) is the area of the  parallelogram spaned by these vectors.

\subsubsection*{2. Cases $N=3$ and $N=4$}

The method of solution of equations (\ref{29d}-\ref{33d}) and equation
(\ref{38d}) is the same for dimensions $N=3$ and $N=4$. So, we can consider
both cases together up to the moment when we shall look for the variables
$\xi_0(t)$, $\xi_1(t)$ and $\xi_2(t)$. Equivalently to (\ref{29d}-\ref{33d}) and
(\ref{38d}) we can consider
\begin{align}
 \label{54d} &\frac{dq_1}{dt}=4(r_k+r_1)p_1,\nonumber\\
&\frac{dp_1}{dt}=-4(r_k+r_1)q_1+4\varrho r_1,\\
 &\frac{dr_1}{dt}=4\varrho (a^2-\lambda_1^2)p_1,\nonumber
\end{align}
with integrals of motion given by
\begin{align}
 \label{55d}
&p_k+p_1=0,\nonumber\\
& q_k+q_1=\varrho,\\
& \varrho(a^2-\lambda_k^2)q_k+\varrho(a^2-\lambda_1^2)q_1-\frac 12(r_k+r_1)^2=g,\nonumber\\
& \frac{r_k}{a^2-\lambda_k^2}+\frac{r_1}{a^2-\lambda_1^2}=f,\nonumber
\end{align}
where $k=0$ for $N=3$ and $k=2$ for $N=4$.

From (\ref{55d}) we express the functions $r_k(t)$, $q_k(t)$ and $q_1(t)$ by $r_1(t)$
\begin{align}
 \label{56d}
& r_k(t)=(a^2-\lambda_k^2)f-\frac{a^2-\lambda_k^2}{a^2-\lambda_1^2}r_1(t),\\
& q_k(t)=\frac{1}{\varrho(\lambda_k^2-\lambda_1^2)}\left(\varrho^2(a^2-\lambda_1^2)-g-\frac 12\left((a^2-\lambda_k^2)f+\frac{\lambda_k^2-\lambda_1^2}{a^2-\lambda_1^2}r_1(t)\right)^2\right),\nonumber\\
& q_1(t)=\frac{1}{\varrho(\lambda_k^2-\lambda_1^2)}\left(-\varrho^2(a^2-\lambda_1^2)+g+\frac 12\left((a^2-\lambda_k^2)f+\frac{\lambda_k^2-\lambda_1^2}{a^2-\lambda_1^2}r_1(t)\right)^2\right)\nonumber
\end{align}
and after substituting to (\ref{54d}) this gives a new integral of motion
\begin{equation}
 \label{57d}
p_1^2-w_4(r_1)=:e=const,
\end{equation}
where
\begin{equation}
 \label{58d}
w_4(r_1)=-\frac{(\lambda_k^2-\lambda_1^2)^2}{4\varrho(a^2-\lambda_1^2)^4}r_1^4
-\frac{(a^2-\lambda_k^2)(\lambda_k^2-\lambda_1^2)}{\varrho^2(a^2-\lambda_1^2)^3}r_1^3-
\end{equation}
$$-\left(\frac{3(a^2-\lambda_k^2)^2f^2}{\varrho^2(a^2-\lambda_1^2)^2}+
\frac{g-\varrho^2(a^2-\lambda_k^2)}{\varrho^2(a^2-\lambda_1^2)^2}-\frac{1}{a^2-\lambda_1^2}\right)r_1^2-
$$$$
-\frac{2(a^2-\lambda_k^2)(g-\varrho^2(a^2-\lambda_k^2)+\frac 12(a^2-\lambda_k^2)^2f^2)f}{\varrho^2(a^2-\lambda_1^2)(\lambda_k^2-\lambda_1^2)}r_1.
$$
Further, substituting (\ref{57d}) into the last equation in (\ref{54d}) we find that
\begin{equation}
 \label{59d}
dt=\pm\frac{dr_1}{4\varrho(a^2-\lambda_1^2)\sqrt{e+w_4(r_1)}},
\end{equation}
i.e. the $r_1(t)$ is elliptic function of the variable $t$.
We have additionally
\begin{align}
 \label{61d}
& p_1(t)=\pm\sqrt{e+w_4(r_1(t))},\\
& p_k(t)=\mp\sqrt{e+w_4(r_1(t))}.
\end{align}
In such a way we find the functions
\begin{align}
 \label{62d}
& \eta_1(t)=e^{i\varphi(t)}(q_1(t)+p_1(t)),\\
& \eta_k(t)=e^{i\varphi(t)}(q_k(t)+p_k(t)),
\end{align}
where
\begin{equation}
 \label{63d}
\varphi(t)=4(a^2-\lambda_k^2)f-4a^2(1+c^2))t+4\frac{\lambda_k^2-\lambda_1^2}{a^2-\lambda_1^2}\int^t_0r_1(s)ds
\end{equation}
and $k=0$ for $N=3$, $k=2$ for $N=4$.

In  case $N=4$ we obtain $\mathbb{R}\ni t\mapsto
\xi_1(t),\xi_2(t)\in\mathbb{C}^2$ using (\ref{17d}-\ref{18dd}) and
(\ref{20d}-\ref{21d}). In  case $N=3$ we obtain $\mathbb{R}\ni t\mapsto
\xi_1(t)\in\mathbb{C}^2$ from (\ref{17d}-\ref{18dd}) and (\ref{23d}-\ref{24d}).
The function $t\mapsto \xi_0(t)\in\mathbb{C}$ is given by
$\xi_0(t)=\sqrt{\eta_0(t)}$.

Summing up, we see that  equation (\ref{3d}) is solved by quadratures  in dimensions $N=2,3$ and $4$. Its solutions are expressed by the elliptic functions and their integrals.

Ending this section let us to express our hope that the equations (\ref{3d}) and
obtained solutions of them find some applications for modeling of nonlinear
phenomena in mechanics and optics, see for example \cite{4}, \cite{6}, \cite{8}.

\section*{Acknowledgements}

The work was partially supported by the Polish grant No. 1 PO3A 001 29.

\bibliographystyle{plain}

\end{document}